\documentclass[%
 reprint,
 amsmath,amssymb,
 aps,
]{revtex4-2}

\usepackage{graphicx}
\usepackage{dcolumn}
\usepackage{bm}
\usepackage[colorlinks,linkcolor=blue,anchorcolor=blue,urlcolor=blue, citecolor=blue]{hyperref}
\usepackage{float}

\newtheorem{theorem}{Theorem}
\begin{document}

\preprint{APS/123-QED}

\title{Security of decoy-state quantum key distribution with correlated intensity fluctuations}

\author{Xoel Sixto$^{1,2}$}
\email{xsixto@com.uvigo.es}
\author{Víctor Zapatero$^{1,2}$}
\email{vzapatero@com.uvigo.es}  
\author{Marcos Curty$^{1,2}$}
\affiliation{$^1$ Escuela de Ingeniería de Telecomunicación, Department of Signal Theory and Communications, ­University of Vigo, Vigo E-36310, Spain
}%

\affiliation{ $^2$ AtlanTTic Research Center, University of Vigo, E-36310, Spain
}%
\date{\today}

\begin{abstract}
One of the most prominent techniques to enhance the performance of practical quantum key distribution (QKD) systems with laser sources is the decoy-state method. Current decoy-state QKD setups operate at GHz repetition rates, a regime where memory effects in the modulators and electronics that control them create correlations between the intensities of the emitted pulses. This translates into information leakage about the selected intensities, which cripples a crucial premise of the decoy-state method, thus invalidating the use of standard security analyses. To overcome this problem, a novel security proof that exploits the Cauchy-Schwarz constraint has been introduced recently. Its main drawback is, however, that the achievable key rate is significantly lower than that of the ideal scenario without intensity correlations. Here, we improve this security proof technique by combining it with a fine-grained decoy-state analysis, which can deliver a tight estimation of the relevant parameters that determine the secret key rate. This results in a notable performance enhancement, being now the attainable distance double than that of previous analyses for certain parameter regimes. Also, we show that when the probability density function of the intensity fluctuations, conditioned on the current and previous intensity choices, is known, our approach provides a key rate very similar to the ideal scenario, which highlights the importance of an accurate experimental characterization of the correlations.
\end{abstract}

\maketitle


\section{Introduction}\label{Introduction}

Quantum key distribution (QKD) offers a way to distribute a secret key over the distance between two communicating parties, Alice and Bob \cite{extra1,extra2,extra3}. When used in conjunction with the one-time-pad encryption scheme~\cite{Vernam}, QKD allows for information-theoretically secure communications, regardless of the future evolution of classical or quantum computers. This is so because its security is based on the laws of quantum mechanics and does not rely on computational assumptions. In recent years, QKD has progressed very rapidly both in theory and in practice, turning into a flourishing commercial technology that is being deployed in metropolitan and intercity fiber-based networks worldwide \cite{nueva1, nueva2, nueva3, nueva4}, including also satellite links~\cite{sat1, sat2, sat2b, sat3}.

One of the most successful QKD protocols for long-distance transmission is undoubtedly the decoy-state QKD scheme~\cite{decoy1, decoy2, Lo_2005}. It uses phase-randomized weak coherent pulses (PRWCPs) emitted by laser sources to provide a secret key rate that scales linearly with the channel transmission~\cite{security_decoy}. Since its first theoretical proposal, numerous experimental implementations have been reported in recent years~\cite{exp_decoy1,exp_decoy2,exp_decoy3,exp_decoy4,exp_decoy5,exp_decoy5b,exp_decoy6,exp_decoy7,exp_decoy8}, being the actual distance record over fiber 421 km~\cite{exp_decoy_long}. Decoy-state QKD has also been demonstrated over satellite links~\cite{sat1, sat2b}, and various companies currently implement it in their commercial products. Importantly, the use of decoy-states is also an essential ingredient for other QKD schemes that use laser sources, like {\it e.g.}, measurement-devide-independent (MDI) QKD~\cite{mdi} or twin-field (TF) QKD~\cite{twinf}, the latter presently holding a distance record over fiber of 833 km~\cite{exp_tf}.

However, despite these tremendous achievements, there are still certain challenges that need to be overcome for QKD to become a widely used technology. One of these challenges is to increase the secret key rate delivered by current experimental setups, which is severely affected by the limited transmissivity of single photons in optical fibers, as well as by the dead-time of the detectors at the receiver. Possible approaches for this include the use of multiplexing techniques---like {\it e.g.} wavelength division multiplexing---that simultaneously transmit several QKD channels over the same fiber~\cite{multi1, multi2}, the adoption of high-dimensional QKD, which can encode many bits of information on a single-photon~\cite{HD_QKD}, and the increase of the pulse repetition rate of the sources~\cite{Fadri}. Indeed, current decoy-state QKD experimental setups operate at GHz repetition rates~\cite{exp_decoy5, exp_decoy7, exp_decoy8, exp_decoy_long}, and the trend is to increase their clock frequency even further. However, in such high-speed regime, memory effects in the modulators and electronics that control them create correlations between the generated optical pulses~\cite{japoneses, Fadri}. That is, the state of a quantum signal emitted by the source at a certain time instant depends not only on the state preparation settings selected by Alice in that time instant, but also on those selected by her in previous time instants. Importantly, if this effect is not properly taken into account in the security proof of QKD, it might open a security loophole in the form of information leakage~\cite{QKD_correlated}. This is so because the settings chosen to encode each quantum signal are also partially leaked through the quantum states of subsequent signals. 

So far, most security proofs of QKD have neglected the effect of pulse correlations and assume independent and identically distributed emitted pulses. Therefore, they cannot be used to guarantee the security of high-speed QKD implementations. Only a few recent works partially address this problem. Precisely, the authors of~\cite{corr1,corr2} studied the case of setting-choice-independent pulse correlations, in which the emitted pulses can be arbitrary correlated between them, but these correlations do not depend on the state preparation setting choices. This scenario may occur, for example, when the temperature of Alice's laser drifts slowly over time due to thermal effects, or when her modulators' power supply fluctuates in time. More recently, various results that address the problem of setting choice-dependent pulse correlations have also been reported. In particular,~\cite{QKD_correlated} introduced a security proof that can handle arbitrary correlations that originate from the phase modulator that encodes the bit and basis information of each generated signal. Likewise, a restricted class of nearest-neighbour pulse correlations that arise from the intensity modulator used to prepare decoy states has been studied in~\cite{japoneses}, where a post-processing technique to treat these correlations is also provided. Also, the authors of~\cite{Zapatero} introduced a general methodology to treat arbitrary intensity correlations in a decoy-state QKD setup. For this, they exploit a fundamental constraint that is a direct consequence of the Cauchy-Schwarz (CS) inequality in Hilbert spaces~\cite{lo,QKD_correlated}. The case of correlations between the global phases of the coherent states emitted by a laser when is operated under gain-switching conditions has been studied in~\cite{corr3}. 

A main limitation of the security proof technique presented in~\cite{Zapatero} is, however, that the delivered secret key rate is significantly lower than that of the ideal scenario without correlations. Only when the intensity correlations are very tiny, the resulting performance can approximate the ideal scenario. In this paper, we improve the general methodology introduced in~\cite{Zapatero}, based on the CS constraint, by combining it with a fine-grained decoy-state analysis. In doing so, we achieve a much tighter estimation procedure to determine the relevant parameters that enter the secret key rate formula. This results in a notably enhancement of the achievable secret key rate in the presence of intensity correlations. Indeed, for certain parameter regimes, the maximum attainable distance can be double than that of~\cite{Zapatero}. In addition, we show that when the probability density function of the intensity fluctuations, conditioned on the current and previous intensity setting choices, is known, our analysis can provide a secret key rate that is very close to that of the ideal scenario. This highlights the importance of properly characterizing intensity correlations in QKD experimental setups. Most importantly, our results can be readily applied to guarantee the security of high-speed decoy-state QKD realizations with current technology, without much penalization on their secret key rate.

The paper is organized as follows. In Sec. \ref{previous} we present the problem of intensity correlations, and explain why the standard decoy-state analysis cannot be directly applied to this scenario. Here, we also emphasize our main contributions in relation to previous results. Next, in Sec. \ref{Initial}, we provide the main assumptions that we consider in the security analysis. Then, in Sec. \ref{Quantifying}, we introduce a fine-grained decoy-state parameter estimation method that can be used in the presence of intensity correlations to tightly estimate the relevant parameters that determine the secret key rate. This estimation procedure is then applied to two different scenarios in the following two sections. Precisely, in Sec.~\ref{Model}, we consider the general case where Alice and Bob only know the intervals where the intensity fluctuations lie. Coming next, the results for the case where they know the probability density function of the intensity fluctuations are presented in Sec.~\ref{Truncated}, followed by the conclusions in  Sec.~\ref{Discussion}. To improve the readability of the paper, certain derivations are omitted in the main text and are included in a series of appendices.

\section{Relation with previous work}\label{previous}

\begin{figure}[H]
\centering\includegraphics [width=6.0cm, height=4cm]{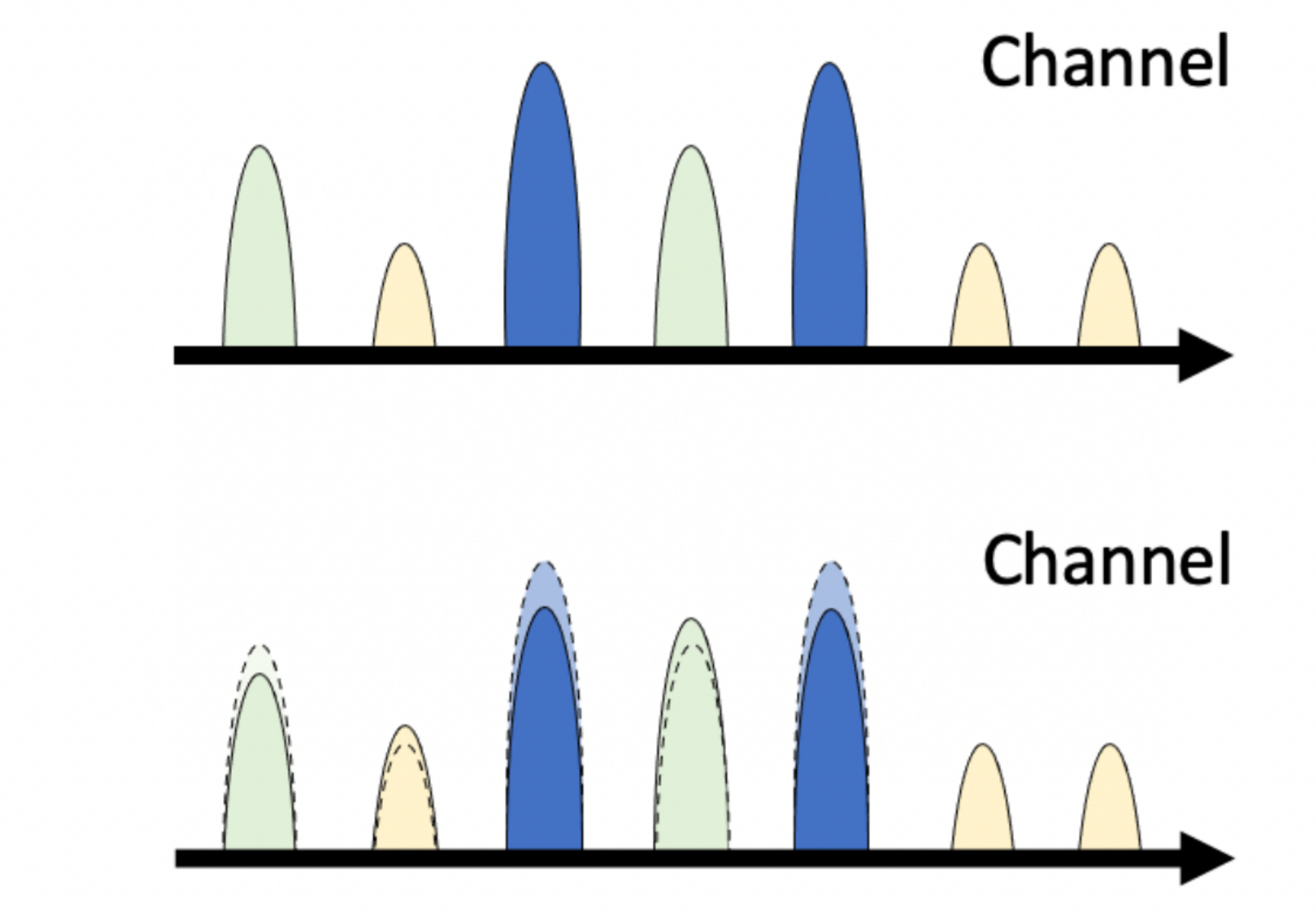}
\caption{\label{fig:correlations_dibujo} Illustration of the effect of intensity correlations. The upper subfigure shows a train of optical pulses sent by Alice to the channel with three different intensity values selected at random, which are indicated with the different colours and amplitudes of the pulses. In the absence of intensity correlations, the intensity of each signal is determined only by the intensity setting selected in the corresponding time instant. In the lower subfigure, we show a specific example of nearest-neighbour intensity correlations. Solid lines indicate the pulses actually emitted, while dashed lines indicate the selected intensity settings. Precisely, this example assumes for illustration purposes that when the intensity setting of the previous pulse is lower (higher) than that of the actual pulse, then the actual intensity generated is a bit lower (higher) than that indicated by the setting selected.
}
\end{figure}

As already mentioned, intensity correlations refer to the fact that the intensity of an optical pulse generated in a certain round of a QKD protocol depends not only on the intensity setting selected for that round, but also on the intensity settings selected for previous rounds. This effect has been experimentally quantified in~\cite{japoneses, Fadri}. It is illustrated in Fig.~\ref{fig:correlations_dibujo} with a simple example. In general, an eavesdropper (Eve) could exploit these correlations to learn information about the intensity settings selected in previous rounds by measuring the intensity of the pulses emitted in subsequent rounds. This could allow her to make the photon-number detection statistics ({\it i.e.}, the yields and error rates associated to $n$-photon pulses) dependent on the intensity setting selected, in so breaking the elementary security premise of the standard decoy-state method~\cite{decoy1, decoy2, Lo_2005}. This situation is similar to that of a Trojan-horse attack~\cite{tha1,tha2,tha3}, in which Eve can learn partial information about the intensities selected to generate the different pulses.

To overcome this problem, there are two main complementary approaches. Precisely, one could implement hardware countermeasures to try to reduce (or even eliminate) the correlations, and one could also include their effect in the security proof of QKD. An example of the first type of approach could be to use various intensity modulators, each of them acting on different non-consecutive signals, such that their modulation rate effectively decreases to a regime where no correlations are created. An example of the second approach has been recently provided in~\cite{japoneses}, where the authors studied a restricted type of nearest-neighbour intensity correlations ({\it i.e.}, this refers to the fact that the actual intensity of each pulse only depends on the intensity settings selected for that time instant and for the previous time instant). In particular,~\cite{japoneses} analyzes the case where only certain intensity settings have a significant effect on the actual intensity of a subsequent pulse, and introduces a post-processing technique to guarantee security in this scenario. In addition, a preliminary experimental characterization of the probability density function of the correlations is provided, which seems to indicate that it essentially follows a gaussian-shaped form, though this knowledge is not exploited by the post-processing technique introduced. 

More recently, in \cite{Zapatero}, the authors presented an asymptotic security proof that can deal with intensity correlations of arbitrary range ({\it i.e.}, not necessarily nearest-neighbour pulse correlations) in the decoy-state parameter estimation procedure. Their method is rather general and can be applied when all previous intensity settings can influence the actual intensity of subsequent pulses. The main idea in~\cite{Zapatero} is to pose a restriction on the maximum bias that Eve can induce between the $n$-photon yields and error rates associated to different intensity settings, by using the CS constraint~\cite{lo,QKD_correlated}. Notably, only two parameters, the correlation range $\xi$ and the maximum deviation $\delta_{\text{max}}$ between the physical intensity and the selected intensity setting, are needed. However, despite the fact that this latter work is quite simple and experimental-friendly, it treats the deviation of each pulse independently of the previous sequence of selected intensities, but instead it takes the maximum possible deviation for the worst-case scenario, thus providing loose bounds. Indeed, the resulting bounds on the secret key rate are significantly lower than those obtained in the absence of correlations. Therefore, the question arises of whether this damaging effect on the key rate is a fundamental feature of the correlations, or rather an artifact of a loose parameter estimation procedure. 

In this paper we preserve the essence of the approach introduced in \cite{Zapatero} to deal with arbitrary intensity correlations---{\it i.e.}, we exploit the CS constraint to quantify the maximum bias that Eve may induce between the photon-number detection statistics associated to different intensity settings---but sharpen the decoy-state method with a finer-grained analysis of the yields and errors that keeps explicit track of the record of settings. This, in turn, allows to incorporate finer-tuned CS constraints to the parameter estimation procedure. Putting it all together, our approach enables a noticeable enhancement of the secret key rate when compared to the results in~\cite{Zapatero}.

In addition, we show that, when the probability density function of the intensity fluctuations is known, our fine-grained decoy-state analysis is rather tight, as it can produce a secret key rate that is comparable to that obtainable in the absence of intensity correlations. This is explicitly illustrated by considering a gaussian-shaped probability density function for the correlations, following the preliminary results in~\cite{japoneses} (see also~\cite{maka}). In this scenario, the principal improvement mainly comes from the fact that the knowledge of the probability density function now permits to calculate certain quantities---that are necessary to determine the secret key rate---precisely, avoiding the need to use looser bounds that exploit monotonicity arguments.

\section{Assumptions}\label{Initial}

For concreteness, below we shall consider a typical polarization encoding decoy-state BB84 protocol with three intensity settings. Nevertheless, our results apply to other encoding schemes, and can be straightforwardly adapted to other decoy-state based QKD protocols with a different number of intensity settings.

In the first place, let us fix the notation that is used throughout the paper. In each round $k$ of the protocol, with $k=1,...,N$, Alice selects an intensity setting $a_{k} \in A=\{\mu, \nu, \omega\}$ with probability $p_{a_{k}}$, a basis $x_{k} \in B=\{\mathrm{X}, \mathrm{Z}\}$ with probability $q_{x_{k}}$ and a uniform raw key bit $r_{k} \in \mathbb{Z}_{2}=\{0,1\}$. Without loss of generality, we impose that the intensity settings satisfy $\mu>\nu>\omega \geq 0$. Then, she encodes the BB84 state defined by $x_{k}$ and $r_{k}$ in a PRWCP with intensity setting $a_{k}$, and sends it to Bob through the quantum channel. Importantly, the actual mean photon-number of the pulse might not match the setting $a_{k}$ due to the presence of intensity correlations. To finish with, we assume perfect phase randomization, perfect polarization encoding, and that there are no side-channels beyond intensity correlations for simplicity.

On the other hand, Bob selects a basis $y_{k}\in B$  with probability $q_{y_{k}}$ and performs a measurement described by a positive operator-valued measure (POVM) $\{\hat{M}_{B_{k}}^{y_{k}, s_{k}}\}_{s_{k} \in\{0,1, f\}}$ on the incident pulse. Here, $B_{k}$ denotes Bob’s $k$-th incoming pulse, $s_{k}$ stands for Bob’s classical outcome and $f$ stands for “no click”. As usual, the basis-independent detection efficiency condition is assumed, \textit{i.e.,} $\hat{M}_{B_{k}}^{\mathrm{Z}, f}=\hat{M}_{B_{k}}^{\mathrm{X}, f}$, and thus we shall simply denote these two operators by $\hat{M}_{B_{k}}^{f}$. This assumption could be removed by the use of MDI-QKD \cite{mdi} or TF-QKD \cite{twinf}.

Let $\vec{a}_{k}= a_{k},a_{k-1},...,a_{1}$ symbolize the record of intensity settings up to round $k$, where $a_{j}\in A$ for every $j$, and let $\alpha_{k}$ stand for the actual intensity emitted in round $k$. We consider that $\alpha_{k}$ is a continuous random variable whose probability density function is fixed by the record of settings  $\vec{a}_{k}$. From now on, we denote such correlation function as $g_{\vec{a_{k}}} (\alpha_{k})$. Three additional elementary assumptions of our work are listed below.

{\it Assumption 1.} The presence of correlations does not compromise the poissonian character of the photon-number statistics of the source conditioned on the value of the actual intensity, $\alpha_{k}$. Mathematically, this amounts to saying that for any given round $k$, and for all $n_{k}\in \mathbb{N}$,
\begin{equation}
p\left(n_{k}| \alpha_{k}\right)=\frac{e^{-\alpha_{k}} \alpha_{k}^{n_{k}}}{n_{k} !}.
\end{equation}
Notably, this assumption is supported by recent high-speed QKD experiments \cite{japoneses, Fadri}. Still, we remark that our analysis could be easily adapted to consider any other photon-number statistics conditioned on the value of $\alpha_{k}$.  
    
{\it Assumption 2.} The intensity correlations have a finite range $\xi$, meaning that the value of the physical intensity of round $k$, $\alpha_{k}$, is not affected by those previous settings $a_{j}$ with $k-j > \xi$. 
    
{\it Assumption 3.} The correlation function $g_{\vec{a_{k}}}(\alpha_{k})$ is only nonzero for $\alpha_{k} \in\left[a_{k}^{-}, a_{k}^{+}\right]$ with  $a_{k}^{\pm}=a_{k}\left(1 \pm \delta^{\pm}_{\vec{a_{k}} }\right)$, where $\delta^{\pm}_{\vec{a_{k}}}$ are the relative deviations. Note that, in virtue of Assumption 2, $\delta^{+}_{\vec{a_{k}} }$ and $\delta^{-}_{\vec{a_{k}} }$ only depend on $a_{k}$ and the previous $\xi$ intensity settings. 

From these three assumptions, it follows that the photon-number statistics for a given round $k$ and a given record of settings $\vec{a}_{k}$ are
\begin{equation}
\label{pns}
\left.p_{n_{k}}\right|_{\vec{a}_{k}}=\int_{a_{k}^{-}}^{a_{k}^{+}} g_{\vec{a_{k}}}(\alpha_{k}) \frac{e^{-\alpha_{k}} \alpha_{k}^{n_{k}}}{n_{k} !} d \alpha_{k}; \quad \text{for all}\quad n_{k} \in \mathbb{N} .
\end{equation}

Note that assumption 2 is not explicitly imposed here.

\section{Quantifying the effect of intensity correlations on the decoy-state parameter estimation procedure}\label{Quantifying}

With the three assumptions stated above and to account for the influence of intensity correlations in the decoy-state analysis, we use the fundamental CS constraint \cite{lo,QKD_correlated}. This result poses a natural constraint between the measurement statistics of two non-orthogonal states. Hence, in particular, one can use it to restrict the possible bias that Eve may induce between the yields (error probabilities) associated to different records of settings. The reader is directed to Appendix \ref{sec:methodsA} for a definition of this statement. The limits we derive with it are shown below.

We define both the yield and the error probability, for any given round $k$, photon-number $n \in\mathbb{N}$ and record of settings $v_{0},...,v_{\xi} \in A$ as:
\begin{equation}
\label{yield}
\begin{aligned}
&Y_{n, v_{0},...,v_{\xi}}^{(k)}=p(s_{k} \neq f|n_{k}=n, a_{k}=v_{0},...,a_{k-\xi}=v_{\xi},\\
&x_{k}=\mathrm{Z}, y_{k}=\mathrm{Z}),\\
&H_{n, v_{0},...,v_{\xi},r}^{(k)}=p(s_{k} \neq f, s_{k} \neq r_{k}|n_{k}=n, a_{k}=v_{0},\\
&...,a_{k-\xi}=v_{\xi}, x_{k}=\mathrm{X}, y_{k}=\mathrm{X}, r_{k}=r).
\end{aligned}
\end{equation}

In virtue of the CS constraint, for any two distinct intensity settings $v_{0}$ and $w_{0}$ that could be selected in round $k$, and for any record of previous settings $v_{1},...,v_{\xi}$, in Appendix \ref{sec:methodsA} it is shown that the associated yields and error probabilities satisfy
\begin{equation}
\label{cs_text}
\begin{aligned}
&G_{-}\left(Y_{n, v_{0}...v_{\xi}}^{(k)}, \tau_{v_{0} w_{0}...v_{\xi}, n}^{\xi}\right) \leq Y_{n, w_{0}...v_{\xi}}^{(k)} \leq\\
&G_{+}\left(Y_{n, v_{0}...v_{\xi}}^{(k)}, \tau_{v_{0} w_{0}...v_{\xi}, n}^{\xi}\right),\\
\end{aligned}
\end{equation}
and
\begin{equation}
\label{cs_text2}
\begin{aligned}
&G_{-}\left(H_{n, v_{0}...v_{\xi}, r}^{(k)}, \tau_{v_{0} w_{0}...v_{\xi}, n}^{\xi}\right) \leq H_{n, w_{0}...v_{\xi}, r}^{(k)} \leq\\ &G_{+}\left(H_{n, v_{0}...v_{\xi}, r}^{(k)}, \tau_{v_{0} w_{0}...v_{\xi}, n}^{\xi}\right),
\end{aligned}
\end{equation}
where
\begin{equation}
\label{eq6}
\begin{aligned}
&G_{-}(y, z)= \begin{cases} g_{-}(y, z) & \text { if } y>1-z \\
0 & \text { otherwise }\end{cases}\\
&G_{+}(y, z)= \begin{cases}g_{+}(y, z) & \text { if } y<z \\
1 & \text { otherwise }\end{cases},
\end{aligned}
\end{equation}
with the function $g_{\pm}(y, z)=y+(1-z)(1-2 y) \pm 2 \sqrt{z}(1-z) y(1-y)$. That is, Eqs.~(\ref{cs_text})-(\ref{cs_text2}), state, respectively, how much $Y_{n, w_{0}...v_{\xi}}^{(k)}$ and $H_{n, w_{0}...v_{\xi}, r}^{(k)}$ can deviate from $Y_{n, v_{0}...v_{\xi}}^{(k)}$ and $H_{n, v_{0}...v_{\xi}, r}^{(k)}$. Crucially, the parameters $\tau_{v_{0} w_{0}...v_{\xi}, n}^{\xi}$ are a lower bound on the squared overlap of the two quantum states with which the two yields are calculated. Explicit expressions for the parameters $\tau_{v_{0} w_{0}...v_{\xi}, n}^{\xi}$ are derived in the Appendix \ref{sec:methodsB} for the two different scenarios that we consider in Secs. \ref{Model} and \ref{Truncated}.

To enable the use of linear programming for the decoy-state parameter estimation procedure, a linear version of the constraints given by Eq.~(\ref{cs_text}) is needed. As shown in Appendix \ref{sec:methodsA}, the linearized versions of the CS constraints for the yields and error probabilities can be expressed as
\begin{widetext}
\begin{equation}
\label{linearyield}
\begin{aligned}
&G_{-}\left(\tilde{Y}_{n,v_{0}...v_{\xi} }^{(k)}, \tau_{v_{0}w_{0}...v_{\xi}, n}^{\xi}\right)+G_{-}^{\prime}\left(\tilde{Y}_{n, v_{0}...v_{\xi}}^{(k)}, \tau_{v_{0}w_{0}...v_{\xi}, n}^{\xi}\right)\left(Y_{n, v_{0}...v_{\xi}}^{(k)}-\tilde{Y}_{n, v_{0}...v_{\xi}}^{(k)}\right) \leq Y_{n, w_{0}...v_{\xi}}^{(k)} \leq \\
&G_{+}\left(\tilde{Y}_{n, v_{0}...v_{\xi}}^{(k)}, \tau_{v_{0}w_{0}...v_{\xi}, n}^{\xi}\right)+G_{+}^{\prime}\left(\tilde{Y}_{n, v_{0}...v_{\xi}}^{(k)}, \tau_{v_{0} w_{0}...v_{\xi}, n}^{\xi}\right)\left(Y_{n, v_{0}...v_{\xi}}^{(k)}-\tilde{Y}_{n, v_{0}...v_{\xi}}^{(k)}\right),
\end{aligned}
\end{equation} 
and
\begin{equation}
\label{linearerror}
\begin{aligned}
&G_{-}\left(\tilde{H}_{n,v_{0}...v_{\xi} }^{(k)}, \tau_{v_{0}w_{0}...v_{\xi}, n}^{\xi}\right)+G_{-}^{\prime}\left(\tilde{H}_{n, v_{0}...v_{\xi}}^{(k)}, \tau_{v_{0}w_{0}...v_{\xi}, n}^{\xi}\right)\left(H_{n, v_{0}...v_{\xi}}^{(k)}-\tilde{H}_{n, v_{0}...v_{\xi}}^{(k)}\right) \leq H_{n, w_{0}...v_{\xi}}^{(k)} \leq \\
&G_{+}\left(\tilde{H}_{n, v_{0}...v_{\xi}}^{(k)}, \tau_{v_{0}w_{0}...v_{\xi}, n}^{\xi}\right)+G_{+}^{\prime}\left(\tilde{H}_{n, v_{0}...v_{\xi}}^{(k)}, \tau_{v_{0} w_{0}...v_{\xi}, n}^{\xi}\right)\left(H_{n, v_{0}...v_{\xi}}^{(k)}-\tilde{H}_{n, v_{0}...v_{\xi}}^{(k)}\right),
\end{aligned}
\end{equation}
\end{widetext}
where both $\tilde{Y}_{n, v_{0}...v_{\xi}}^{(k)}$ and $\tilde{H}_{n, v_{0}...v_{\xi}}^{(k)}$ are the reference parameters of the linear approximations, introduced in Appendix \ref{appendix:values}. Finally, the functions $G_{\pm}^{\prime}$ are defined as
\begin{equation}\label{brussels}
\begin{aligned}
&G_{-}^{\prime}(y, z)=\begin{cases}g_{-}^{\prime}(y, z) & \text { if } y>1-z \\
0 & \text { otherwise }\end{cases}
\\
&G_{+}^{\prime}(y, z)= \begin{cases}g_{+}^{\prime}(y, z) & \text { if } y<z \\
0 & \text { otherwise }\end{cases},
\end{aligned}
\end{equation}
with $g_{\pm}^{\prime}(y, z)=-1+2 z \pm(1-2 y) \sqrt{z(1-z) / y(1-y)}$.

\section{Model-independent correlations}\label{Model}

\subsection{Characterization}\label{Characterization}

In this section we now consider a general scenario in which the correlation function $g_{\vec{a}_{k}}(\alpha_{k})$ given by Eq.~\eqref{pns} is unknown. This implies that one cannot compute the photon-number statistics of the emitted pulses explicitly, and thus must impose bounds on them by invoking monotonicity arguments. We consider two cases.

In the first one, the maximum relative deviations  $\delta_{(a_{k} \ldots a_{k-\xi})}^{+}$ and $\delta_{(a_{k} \ldots a_{k-\xi})}^{-}$ depend on the intensity setting choices corresponding to all the previous $\xi$ rounds and the present round, and define an interval that is in general not symmetric with respect to the value of the selected setting. As shown in Appendix \ref{sec:methodsB}, and denoting $a_{1}^{N}=a_{1} \ldots a_{N}$, the bound for this case reads
\begin{widetext}
\begin{equation}
\label{cotavariable_texto}
\begin{aligned}
\sqrt{\tau_{v_{0},w_{0},...,v_{\xi}, n}^{\text{$\xi$}}} =
\sum_{a_{k+1}^{\min\{k+\xi,N\}}}\prod_{i=k+1}^{\min\{k+\xi,N\}} p_{a_{i}}  \left[e^{\frac{1}{2}(-a_{i}^{(w_{0})+}-a_{i}^{(v_{0})+})}+e^{\frac{1}{2}(-a_{i}^{(w_{0})-}-a_{i}^{(v_{0})-})}\left( e^{\sqrt{a_{i}^{(w_{0})-}a_{i}^{(v_{0})-}}} -1\right)\right],
\end{aligned}
\end{equation}
\end{widetext}
where the terms $a_{i}^{(v_{0}) \pm}$ satisfy $a_{i}^{(v_{0}) \pm}=a_{i}(1 \pm \delta_{\left(a_{i} \ldots a_{k+1},a_{k}=v_{0},a_{k-1}=v_{1} \ldots a_{i-\xi}=v_{\xi+k-i}\right)}^{\pm})$ and analogously for $a_{i}^{(w_{0}) \pm}$. Note that, with this notation, settings $a_{1}$ to $a_{k}$ are fixed to a certain value, while settings $a_{k+1}$ to $a_{i}$ are not. Also, note that the bound given by Eq.~(\ref{cotavariable_texto}) does not depend on the photon-number $n$, {\it i.e.} it holds for all $n$.

Secondly, we consider as well a simplified situation where only a worst-case deviation $\delta_{\text{max}}$ is known for every possible record of settings. That is, it holds that $\delta_{\text{max}}\geq \delta^{\pm}_{(a_{i}...a_{k+1},a_{k}=w_{0},v_{0},a_{k-1}=v_{1}...a_{i-\xi}=v_{\xi+k-i})}$ for all $k=1,...,N$ and all $i\in[k+1, \min\{k+\xi,N\}]$. In this case, the bound simply reads
\begin{equation}
\label{cota_texto}
\begin{aligned}
\sqrt{\tau_{v_{0},w_{0},...,v_{\xi}, n}^{\text{$\xi$}}}= \left[1-\sum_{a_{i}\in A}p_{a_{i}}\left( e^{-a_{i}^{-}}-e^{-a_{i}^{+}} \right)\right]^{\xi} ,
\end{aligned}
\end{equation}
where $a_{i}^{\pm}=a_{i}(1\pm\delta_{\text{max}})$. This derivation is also presented in Appendix \ref{sec:methodsB}. Note that Eq.~\eqref{cotavariable_texto} can only be used if one is able to experimentally characterize the maximum relative deviation for every possible combination of settings.

From now on, and in order to keep the discussion simple, we focus on the case of nearest-neighbour intensity correlations. That is, we set $\xi=1$. For this purpose, it is convenient to do a slight change of notation, by calling $v_{0}=a$; $w_{0}=b$ and $v_{1}=c$ as this makes the following section easier to follow. Note that the analysis below can be easily adapted for the case of $\xi>1$, and we include simulations for this latter scenario in Sec.~\ref{Simulations}.

\subsection{Decoy-state method}\label{Decoy}

To introduce the linear programs that perform the parameter estimation, we shall now provide a decoy-state analysis.

Let us start by defining the $Z$ basis gain for a certain pair of  intensity settings $a$ and $c$, and a number of rounds $N$, as 
\begin{equation}
Z_{a, c, N}=\sum_{k=1}^{N} Z_{a, c}^{(k)}, 
\end{equation}
with $Z_{a,c}^{(k)}=\mathbb{I}_{\left\{a_{k}=a, a_{k-1}=c, x_{k}=y_{k}=\mathrm{Z}, s_{k} \neq f\right\}}$. That is, $Z_{a,c}^{(k)}=1$ if in round $k$, Alice selects an intensity setting $a$ that is preceded by an intensity setting $c$ (in round $k-1$), both Alice and Bob select the $Z$ basis, and a click occurs at Bob's side. Thus:
\begin{eqnarray}\label{LP1}
\left\langle Z_{a, c}^{(k)}\right\rangle&=&p^{(k)}(a, c, \mathrm{Z}, \mathrm{Z}, \text {click})\nonumber \\
&=&q_{\mathrm{Z}}^{2} p_{a} p_{c} \sum_{n=0}^{\infty} p^{(k)}(n,\text{click}|a, c, \mathrm{Z}, \mathrm{Z})\nonumber \\
&=&q_{\mathrm{Z}}^{2} p_{a} p_{c} \sum_{n=0}^{\infty} p^{(k)}(n|a, c) Y_{n, a, c}^{(k)}.
\end{eqnarray}

Straightforward monotonicity arguments lead to the following bounds on the photon-number statistics $p^{(k)}(n|a, c)$:
\begin{equation}
\label{acotando}
\begin{aligned}
&p^{(k)}(0|a, c) \in\left[e^{-a_{c}^{+}}, e^{-a_{c}^{-}}\right],\\
&p^{(k)}(n|a, c) \in\left[\frac{e^{-a_{c}^{-}} (a_{c}^{-})^{n}}{n !}, \frac{e^{-a_{c}^{+}} (a_{c}^{+})^{n}}{n !}\right](n \geq 1),
\end{aligned}
\end{equation}
where $a_{c}^{\pm}=a(1\pm \delta^{\pm}_{(a_{k}=a, a_{k-1}=c)})$. Note that, despite the fact that this notation is similar to that used in Eq.~\eqref{cotavariable_texto}, the parameters $a_{c}^{\pm}$ are actually not equal to $a_{i}^{(\gamma)\pm}$ with $\gamma\in{}A$. This is so because here, only the settings $a_{k}$ and $a_{k-1}$ matter, and they are respectively fixed to $a$ and $c$. 

Now, using these intervals in Eq.~\eqref{LP1}, one obtains
\begin{equation}
\begin{aligned}
&\frac{\left\langle Z_{a, c}^{(k)}\right\rangle}{q_{Z}^{2} p_{a} p_{c}} \geq e^{-a_{c}^{+}} Y_{0, a, c}^{(k)}+\sum_{n=1}^{\infty} \frac{e^{-a_{c}^{-}} (a_{c}^{-})^{n}}{n !} Y_{n, a, c}^{(k)},\\
&\frac{\left\langle Z_{a, c}^{(k)}\right\rangle}{q_{Z}^{2} p_{a} p_{c}} \leq e^{-a_{c}^{-}} Y_{0, a, c}^{(k)}+\sum_{n=1}^{\infty} \frac{e^{-a_{c}^{+}} (a_{c}^{+})^{n}}{n !} Y_{n, a, c}^{(k)},
\end{aligned}
\end{equation}
for all $a,c\in A$ and $k=1,...,N$. Selecting a threshold photon-number for the numerics, $n_{\text{cut}}$, and using the fact that 
\begin{equation}
\sum_{n=n_{\text {cut }}+1}^{\infty} \frac{Y_{n, a, c}^{(k)} e^{-a_{c}^{+}} a_{c}^{+n}}{n!} \leq 1-\sum_{n=0}^{n_{\text {cut }}} \frac{e^{-a_{c}^{+}} (a_{c}^{+})^{n}}{n!},
\end{equation}
we have that
\begin{eqnarray}
\label{LP2}
\frac{\left\langle Z_{a, c}^{(k)}\right\rangle}{q_{\mathrm{Z}}^{2} p_{a} p_{c}} &\geq& e^{-a_{c}^{+}} Y_{0, a, c}^{(k)}+\sum_{n=1}^{n_{\mathrm{cut}}} \frac{e^{-a_{c}^{-}} (a_{c}^{-})^{n}}{n !} Y_{n, a, c}^{(k)}\nonumber\\
\frac{\left\langle Z_{a, c}^{(k)}\right\rangle}{q_{\mathrm{Z}}^{2} p_{a} p_{c}} &\leq& 1-e^{-a_{c}^{+}}+e^{-a_{c}^{-}} Y_{0, a, c}^{(k)}-\sum_{n=1}^{n_{\mathrm{cut}}} \frac{e^{-a_{c}^{+}} (a_{c}^{+})^{n}}{n !}\nonumber \\
&\times&\left(1-Y_{n, a, c}^{(k)}\right).
\end{eqnarray}

Notice how replacing $Z$ by $X$ everywhere, one obtains the corresponding analysis for the $X$ basis gains in a certain round $k$.

We now have to impose similar constraints to the error counts. For that matter, we define the number of $X$ basis error counts with settings $a$ in round $k$ and $c$ in round $k-1$ as 
\begin{equation}
E_{a, c, N}=\sum_{k=1}^{N} E_{a, c}^{(k)}, 
\end{equation}
with $E_{a, c}^{(k)}=X_{a, c}^{(k)} \mathbb{I}_{\left\{r_{k} \neq s_{k}\right\}}$. Then, we have that
\begin{eqnarray}
\left\langle E_{a, c}^{(k)}\right\rangle&=&p^{(k)}(a, c, \mathrm{X}, \mathrm{X}, \mathrm{err})\nonumber\\
&=&q_{\mathrm{X}}^{2} p_{a} p_{c} \sum_{n=0}^{\infty} p^{(k)}(n, \operatorname{err}|a, c, \mathrm{X}, \mathrm{X})\nonumber\\
&=&q_{\mathrm{X}}^{2} p_{a} p_{c} \sum_{n=0}^{\infty} p^{(k)}(n|a, c) H_{n, a, c}^{(k)},
\end{eqnarray}
where we have defined 
\begin{eqnarray}
H_{n, a, c}^{(k)}&=&p^{(k)}(\operatorname{err} \mid n, a, c, \mathrm{X}, \mathrm{X})\nonumber \\
&=&\frac{H_{n, a, c, 0}^{(k)}+H_{n, a, c, 1}^{(k)}}{2}. 
\end{eqnarray}

With the same steps as before, it follows that
\begin{eqnarray}
\label{LP3}
\frac{\left\langle E_{a, c}^{(k)}\right\rangle}{q_{\mathrm{X}}^{2} p_{a} p_{c}} &\geq& e^{-a_{c}^{+}} H_{0, a, c}^{(k)}+\sum_{n=1}^{n_{\text {cut }}} \frac{e^{-a_{c}^{-}} (a_{c}^{-})^{n}}{n !} H_{n, a, c}^{(k)},\nonumber\\
\frac{\left\langle E_{a, c}^{(k)}\right\rangle}{q_{\mathrm{X}}^{2} p_{a} p_{c}} &\leq& 1-e^{-a_{c}^{+}}+e^{-a_{c}^{-}} H_{0, a, c}^{(k)}\nonumber\\
&-&\sum_{n=1}^{n_{\text {cut }}} \frac{e^{-a_{c}^{+}} (a_{c}^{+})^{n}}{n !}\left(1-H_{n, a, c}^{(k)}\right).
\end{eqnarray}

Now, summing over $k$ and dividing by $N$ in both Eq.~\eqref{LP2} and Eq.~\eqref{LP3}, one obtains  bounds for the average parameters 
\begin{eqnarray}
y_{n, a, c, N}&=&\sum_{k=1}^{N} \frac{Y_{n, a, c}^{(k)}}{N}, \nonumber \\
h_{n, a, c, N}&=&\sum_{k=1}^{N} \frac{H_{n, a, c}^{(k)}}{N}, 
\end{eqnarray}
from the round dependent bounds. Thus, defining 
\begin{eqnarray}
\bar{Z}_{a, c, N}&=&\frac{Z_{a, c, N}}{N}, \nonumber \\ 
\bar{E}_{a, c, N}&=&\frac{E_{a, c, N}}{N}, 
\end{eqnarray}
we obtain that the final bounds are:
\begin{eqnarray}
\frac{\left\langle\bar{Z}_{a, c, N}\right\rangle}{q_{\mathrm{Z}}^{2} p_{a} p_{c}} &\geq& e^{-a_{c}^{+}} y_{0, a, c, N}+\sum_{n=1}^{n_{\text {cut }}} \frac{e^{-a_{c}^{-}} (a_{c}^{-})^{n}}{n !} y_{n, a, c, N},\nonumber \\
\frac{\left\langle\bar{Z}_{a, c, N}\right\rangle}{q_{\mathrm{Z}}^{2} p_{a} p_{c}} &\leq& 1-e^{-a_{c}^{+}}+e^{-a_{c}^{-}} y_{0, a, c, N}-\sum_{n=1}^{n_{\text {cut }}} \frac{e^{-a_{c}^{+}} (a_{c}^{+})^{n}}{n !}\nonumber\\
&\times&\left(1-y_{n, a, c, N}\right),\nonumber \\
\frac{\left\langle\bar{E}_{a, c, N}\right\rangle}{q_{\mathrm{X}}^{2} p_{a} p_{c}} &\geq& e^{-a_{c}^{+}} h_{0, a, c, N}+\sum_{n=1}^{n_{\text {cut }}} \frac{e^{-a_{c}^{-}} (a_{c}^{-})^{n}}{n !} h_{n, a, c, N},\nonumber \\ 
\frac{\left\langle\bar{E}_{a, c, N}\right\rangle}{q_{\mathrm{X}}^{2} p_{a} p_{c}} &\leq& 1-e^{-a_{c}^{+}}+e^{-a_{c}^{-}} h_{0, a, c, N}-\sum_{n=1}^{n_{\text {cut }}} \frac{e^{-a_{c}^{+}} (a_{c}^{+})^{n}}{n !}\nonumber\\
&\times&\left(1-h_{n, a, c, N}\right).
\end{eqnarray}

\subsection{Linear programs for parameter estimation}\label{Linear}

In this section, we present the linear programs that allow to estimate the relevant single-photon parameters by putting together the decoy-state constraints, already introduced above, and the linearized CS constraints with the notation presented in Eq.~\eqref{esesta}.

To begin with, note that the quantities that go into the secret key rate are the average number of single-photon counts associated to the case where Alice selects the signal intensity setting, which we shall denote by $\bar{Z}_{1,\mu,N}$, and the average number of phase errors associated to these single-photon counts (which in the case of the BB84 protocol, without state preparation flaws and assuming the asymptotic limit, match the average number of single-photon error counts in the $X$ basis). Moreover, since below we shall consider the asymptotic secret key regime where $p_\mu\approx{}1$, for simplicity we can restrict our estimation to the number of single-photon error counts in the $X$ basis when Alice selects the intensity setting, which we shall denote by $\bar{E}_{1,\mu,N}$.

Let us consider $\bar{Z}_{1,\mu,N}$ first. Formally, 
\begin{equation}
\bar{Z}_{1, \mu, N}=\sum_{k=1}^{N} \frac{Z_{1, \mu}^{(k)}}{N}, 
\end{equation}
where for each $k$, $Z_{1, \mu}^{(k)}=Z_{\mu}^{(k)}\mathbb{I}_{{\{n_{k}=1}\}}$ is the probability that Alice selects the signal intensity setting $\mu$, both Alice and Bob select the $Z$ basis, and a click occurs at Bob's side. In turn, it is clear that $Z_{1,\mu}^{(k)}$ decomposes as 
\begin{equation}
Z_{1,\mu}^{(k)}=\sum_{h\in A} Z_{1,\mu,h}^{(k)}, 
\end{equation}
where of course $Z_{1,\mu,h}^{(k)}= Z_{1,\mu}^{(k)}\mathbb{I}_{{\{a_{k-1}=h}\}}$. Then it trivially follows that:
\begin{eqnarray}
\langle \bar{Z}_{1, \mu, N}\rangle&=&\frac{1}{N}\sum_{k=1}^{N} \langle Z_{1, \mu}^{(k)}\rangle=\frac{1}{N}\sum_{k=1}^{N}\sum_{h\in A} \langle Z_{1, \mu, h}^{(k)}\rangle \nonumber\\
&=&\frac{1}{N}\sum_{k=1}^{N}\sum_{h\in A}q_{\mathrm{Z}}^{2} p_{\mu}p_{h}p^{(k)}(1|\mu, h) Y_{1, \mu, h}^{(k)}\nonumber \\
&\geq&\frac{1}{N}\sum_{k=1}^{N}\sum_{h\in A}q_{\mathrm{Z}}^{2} p_{\mu}p_{h}\mu_{h}^{-} e^{-\mu_{h}^{-}} Y_{1, \mu, h}^{(k)}\nonumber\\
&=&\sum_{h\in A}q_{\mathrm{Z}}^{2} p_{\mu}p_{h}\mu_{h}^{-} e^{-\mu_{h}^{-}} y_{1, \mu, h, N}.
\end{eqnarray}

Note that the above equation lower bounds the expected value of the quantity of interest $\bar{Z}_{1, \mu, N}$ using the yields $y_{1,a,b,N}$ introduced in the previous section. Thus, the linear program of interest is
\begin{widetext}
\begin{gather}\label{LP_final}
\begin{aligned}
\textup{min}\hspace{.3cm}& q_{\mathrm{Z}}^{2}p_{\mu}\sum_{h\in A}p_{h}\mu_{h}^{-}e^{-\mu_{h}^{-}}y_{1,\mu,h,N}\\
\textup{s.t.}\hspace{.3cm}&\frac{\left\langle\bar{Z}_{a, c, N}\right\rangle}{q_{\mathrm{Z}}^{2} p_{a} p_{c}} \geq e^{-a_{c}^{+}} y_{0, a, c, N}+\sum_{n=1}^{n_{\mathrm{cut}}} \frac{e^{-a_{c}^{-}} (a_{c}^{-})^{n}}{n !} y_{n, a, c, N}\quad(a, c \in A), \\
&\frac{\left\langle\bar{Z}_{a, c, N}\right\rangle}{q_{\mathrm{Z}}^{2} p_{a} p_{c}} \leq 1-e^{-a_{c}^{+}}+e^{-a_{c}^{-}} y_{0, a, c, N}-\sum_{n=1}^{n_{\mathrm{cut}}} \frac{e^{-a_{c}^{+}} (a_{c}^{+})^{n}}{n !}\left(1-y_{n, a, c, N}\right) \quad(a, c \in A), \\
&c_{a b c, n}^{+}+m_{a b c, n}^{+} y_{n, a, c, N} \geq y_{n, b, c, N}\left(a, b, c \in A, b \neq a, n=0, \ldots, n_{\mathrm{cut}}\right), \\
&c_{a b c, n}^{-}+m_{a b c, n}^{-} y_{n, a, c, N} \leq y_{n, b, c, N}\left(a, b, c \in A, b \neq a, n=0, \ldots, n_{\mathrm{cut}}\right), \\
&0 \leq y_{n, a, b, N} \leq 1\left(a, b \in A, n=0, \ldots, n_{\mathrm{cut}}\right),
\end{aligned}
\end{gather}
\end{widetext}
with the parameters $c_{a b c, n}^{\pm}$ and $m_{a b c, n}^{\pm}$ that arise from the linear CS constraints given in Appendix~\ref{sec:methodsA}. We shall denote the lower bound on $\langle \bar{Z}_{1, \mu, N}\rangle$ obtained with the linear program above by $\bar{Z}_{1,\mu,N}^{\mathrm{L}}$.

Naturally, substituting $Z$ for $X$ everywhere yields the equivalent program for the average number of signal-setting single-photon counts in the $X$ basis.

Proceeding analogously for the average number of signal-setting single-photon error counts in the $X$ basis we have that 
\begin{equation}
\bar{E}_{1, \mu, N}=\sum_{k=1}^{N} \frac{E_{1, \mu}^{(k)}}{N},
\end{equation}
where for each round $k$, $E_{1, \mu}^{(k)}=E_{\mu}^{(k)}\mathbb{I}_{{\{n_{k}=1}\}}$. As before, $E_{1,\mu}^{(k)}$ decomposes as 
\begin{equation}
E_{1,\mu}^{(k)}=\sum_{h\in A} E_{1,\mu,h}^{(k)}, 
\end{equation}
where $E_{1,\mu,h}^{(k)}= E_{1,\mu}^{(k)}\mathbb{I}_{{\{a_{k-1}=h}\}}$. We obtain therefore that:
\begin{eqnarray}
\langle \bar{E}_{1, \mu, N}\rangle&=&\frac{1}{N}\sum_{k=1}^{N} \langle E_{1, \mu}^{(k)}\rangle=\frac{1}{N}\sum_{k=1}^{N}\sum_{h\in A} \langle E_{1, \mu, h}^{(k)}\rangle \nonumber\\
&=&\frac{1}{N}\sum_{k=1}^{N}\sum_{h\in A}q_{\mathrm{X}}^{2} p_{\mu}p_{h}p^{(k)}(1|\mu, h) H_{1, \mu, h}^{(k)}\nonumber\\ 
&\leq&\frac{1}{N}\sum_{k=1}^{N}\sum_{h\in A}q_{\mathrm{X}}^{2} p_{\mu}p_{h}\mu_{h}^{+} e^{-\mu_{h}^{+}} H_{1, \mu, h}^{(k)}\nonumber\\
&=&\sum_{h\in A}q_{\mathrm{X}}^{2} p_{\mu}p_{h}\mu_{h}^{+} e^{-\mu_{h}^{+}} h_{1, \mu, h, N}.
\end{eqnarray}

Then an upper bound $\bar{E}_{1, \mu, N}^{\mathrm{U}}$ on the quantity $\langle \bar{E}_{1, \mu, N}\rangle$ is achieved by using the following linear program
\begin{widetext}
\begin{gather}
\begin{aligned}
\textup{max}\hspace{.3cm}&  q_{\mathrm{X}}^{2} p_{\mu}\sum_{h\in A} p_{h}\mu_{h}^{+} e^{-\mu_{h}^{+}} h_{1, \mu, h, N}\\
\textup{s.t.}\hspace{.3cm}& \frac{\left\langle\bar{E}_{a, c, N}\right\rangle}{q_{\mathrm{X}}^{2} p_{a} p_{c}} \geq e^{-a_{c}^{+}} h_{0, a, c, N}+\sum_{n=1}^{n_{\mathrm{cut}}} \frac{e^{-a_{c}^{-}} (a_{c}^{-})^{n}}{n !} h_{n, a, c, N}\quad(a, c \in A), \\
&\frac{\left\langle\bar{E}_{a, c, N}\right\rangle}{q_{\mathrm{X}}^{2} p_{a} p_{c}} \leq 1-e^{-a_{c}^{+}}+e^{-a_{c}^{-}} h_{0, a, c, N}-\sum_{n=1}^{n_{\mathrm{cut}}} \frac{e^{-a_{c}^{+}} (a_{c}^{+})^{n}}{n !}\left(1-h_{n, a, c, N}\right) \quad(a, c \in A), \\
&t_{a b c, n}^{+}+s_{a b c, n}^{+} h_{n, a, c, N} \geq h_{n, b, c, N}\left(a, b, c \in A, b \neq a, n=0, \ldots, n_{\mathrm{cut}}\right), \\
&t_{a b c, n}^{-}+s_{a b c, n}^{-} h_{n, a, c, N} \leq h_{n, b, c, N}\left(a, b, c \in A, b \neq a, n=0, \ldots, n_{\mathrm{cut}}\right), \\
&0 \leq h_{n, a, b, N} \leq 1\left(a, b \in A, n=0, \ldots, n_{\mathrm{cut}}\right),
\end{aligned}
\end{gather}
\end{widetext}
with the parameters $t_{a b c, n}^{\pm}$ and $s_{a b c, n}^{\pm}$ that arise from the linear CS constraints given in Appendix~\ref{sec:methodsA}.

\subsection{Simulations}\label{Simulations}

As shown in \cite{Zapatero}, the asympotitc secret key rate is well approximated by
\begin{equation}
\label{skr}
K_{\infty}=\bar{Z}_{1, \mu, N}^{\mathrm{L}}\left[1-h\left(\frac{\bar{E}_{1, \mu, N}^{\mathrm{U}}}{\bar{X}_{1, \mu, N}^{\mathrm{L}}}\right)\right]-f_{\mathrm{EC}} \bar{Z}_{\mu, N} h\left(E_{\mathrm{tol}}\right),
\end{equation}
for large enough N, as long as the variances of the experimental averages vanish asymptotically (see \cite{Zapatero} for a precise meaning of this statement). Here, $h(x)$ denotes the binary entropy function, $f_{\mathrm{EC}}$ is the error correction efficiency, the parameter $\bar{Z}_{\mu, N}$ is the gain in the $Z$ basis defined as
\begin{equation}
\bar{Z}_{\mu, N}=\sum_{h\in A} \bar{Z}_{\mu, h, N},
\end{equation}
and $E_{\mathrm{tol}}$ is the overall error rate observed in the $Z$ basis. 

Note that here one cannot simply use the asymptotic secret key rate formula against collective attacks due to the presence of intensity correlations. To be precise, we have that the asymptotic equivalence between the collective and the coherent settings is typically established on the basis of the so-called post-selection technique \cite{extra4} built on the De Finetti theorem \cite{Finetti, extra5}. However, the round-exchangeability property required to apply the post-selection technique is generally invalidated by pulse-correlations.

Even though simulations with real data from~\cite{japoneses} are presented in Sec.~\ref{Truncated}, in order to compare our fine-grained analysis with the previous results in \cite{Zapatero}, we fix the experimental inputs of the linear programs $\bar{Z}_{a,c, N} / q_{\mathrm{Z}}^{2} p_{a}p_{c}$, $\bar{X}_{a, c, N} / q_{\mathrm{X}}^{2} p_{a}p_{c}$ and $\bar{E}_{a, c, N} / q_{\mathrm{X}}^{2} p_{a}p_{c}$ to their expected values according to a typical channel model, which is provided in Appendix \ref{appendix:values}. 

For that matter, let $\eta_{\mathrm{det}}$ denote the common detection efficiency of Bob’s detectors, and let $\eta_{\mathrm{ch}}=10^{-\alpha_{\text{att}} L / 10}$ be the transmittance of the quantum channel, where $\alpha_{\text{att}}$ (dB/km) is its attenuation coefficient and $L$ (km) is the distance. Also, let $p_{\mathrm{d}}$ denote the dark count probability of each of Bob’s detectors and let $\delta_{\mathrm{A}}$ stand for the polarization misalignment occurring in the channel. This model yields the following results (see Appendix \ref{appendix:values})
\begin{equation}
\frac{\left\langle\bar{Z}_{a, c, N}\right\rangle}{q_{\mathrm{Z}}^{2} p_{a} p_{c}}=\frac{\left\langle\bar{X}_{a, c, N}\right\rangle}{q_{\mathrm{X}}^{2} p_{a} p_{c}}=1-\left(1-p_{\mathrm{d}}\right)^{2} e^{-\eta a}, 
\end{equation}
and
\begin{eqnarray}
\frac{\left\langle\bar{E}_{a, c, N}\right\rangle}{q_{X}^{2} p_{a} p_{c}}&=&\frac{\left\langle\bar{E}_{a, c, N(\mathrm{Z})}\right\rangle}{q_{Z}^{2} p_{a} p_{c}}=\frac{p_{\mathrm{d}}^{2}}{2}+p_{\mathrm{d}}\left(1-p_{\mathrm{d}}\right)\nonumber\\
&\times&\left(1+h_{\eta, a, c, \delta_{\mathrm{A}}}\right)+\left(1-p_{\mathrm{d}}\right)^{2}\nonumber\\
&\times&\left(\frac{1}{2}+h_{\eta, a, c, \delta_{\mathrm{A}}}-\frac{1}{2} e^{-\eta a}\right),
\end{eqnarray}
for $a,c\in A$ and where $\eta=\eta_{\mathrm{det}} \eta_{\mathrm{ch}}$. Here, we define the parameter $h_{\eta, a, c, \delta_{\mathrm{A}}}$ as
\begin{equation}
h_{\eta, a, c, \delta_{\mathrm{A}}}=\frac{e^{-\eta a \cos ^{2} \delta_{\mathrm{A}}}-e^{-\eta a \sin ^{2} \delta_{\mathrm{A}}}}{2}.
\end{equation}  

We also introduce the parameter $\bar{E}_{a, c, N(\mathrm{Z})}$, which is equivalent to $\bar{E}_{a, c, N}$ but accounts for the $Z$ basis error clicks. The tolerated bit error rate of the sifted key is set to $E_{\mathrm{tol}}=\left\langle\bar{E}_{\mu, N(\mathrm{Z})}\right\rangle /\left\langle\bar{Z}_{\mu, N}\right\rangle$. Also, the reference parameters for the linearized CS constraints are fixed by the channel model as well, as indicated in Appendix \ref{appendix:values}.
\begin{figure}[H]
\centering\includegraphics [width= 8.6cm, height=6cm] {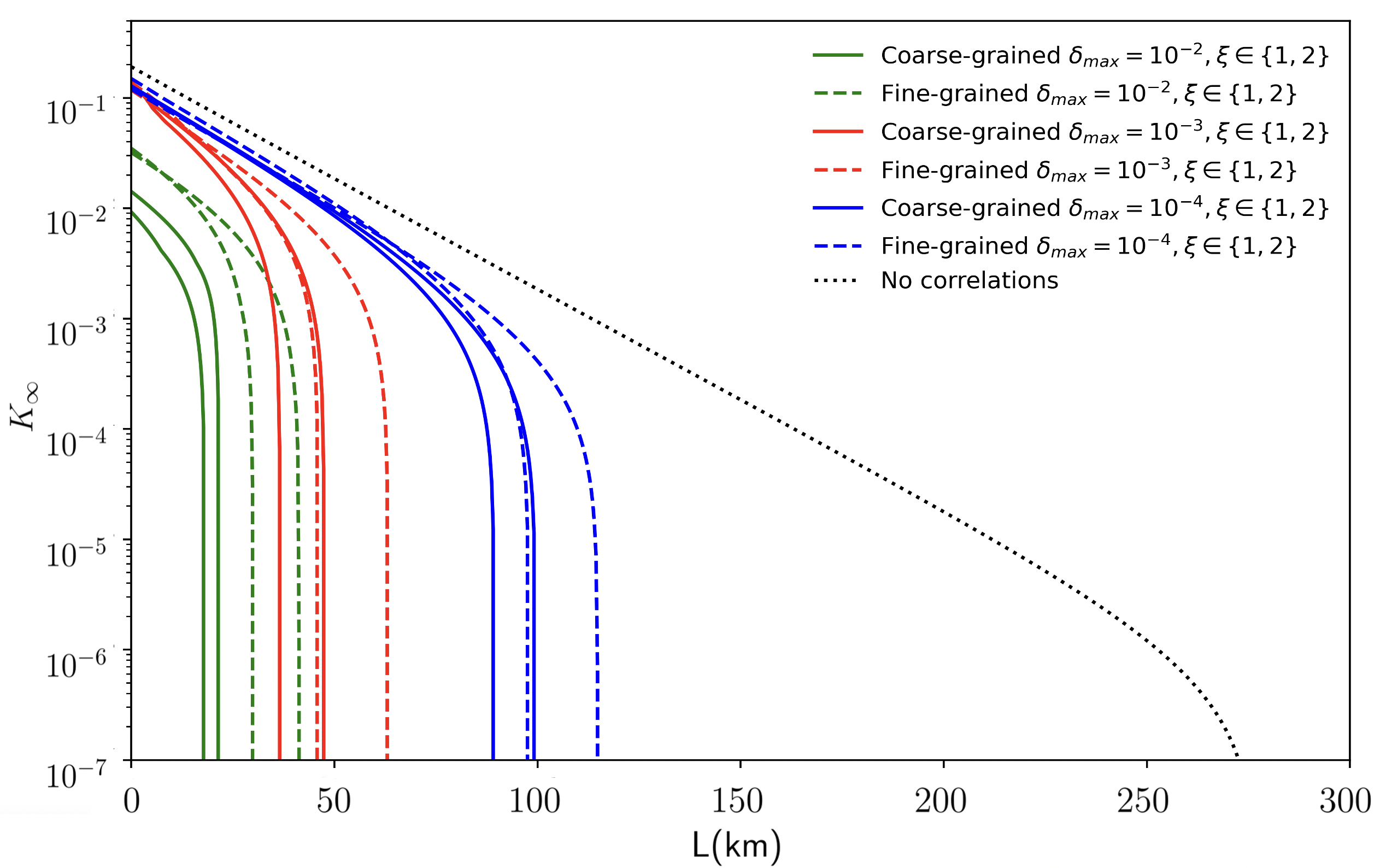}
\caption{\label{fig:fixed} Secret key rate given by Eq.\eqref{skr} using the analysis provided by Eq.~\eqref{cota_texto}. We consider different values for the maximum deviation $\delta_{\text{max}}\in\{10^{-2},10^{-3},10^{-4}\}$ between the actual physical intensity $\alpha_{k}$ and the selected intensity setting $a_{k}$, and two values of the parameter $\xi$. Precisely, we consider $\xi=1$ ({\it i.e.} corresponding to the case of nearest-neighbour pulse correlations) and $\xi=2$ (corresponding to pulse correlations of range two). The coarse-grained analysis follows the techniques presented in~\cite{Zapatero}. 
}
\end{figure}

In particular, in the simulations we take $\eta_{\mathrm{det}}=0.65$ and $p_{\mathrm{d}}=7.2 \times 10^{-8}$ \cite{decrease1}. The attenuation coefficient of the channel is set to $\alpha_{\mathrm{att}}=0.2\text{ dB/km}$, the error correction efficiency to $f_{EC}=1.16$ and a channel misalignment of $\delta_{A}=0.08$ is used. As for the intensities, the weakest intensity setting is set to $\omega=10^{-4}$ due to the finite extinction ratio of intensity modulators. For simplicity, due to the large number of constraints included in the linear programs, instead of optimizing both $\mu$ and $\nu$ to maximize the asymptotic secret key rate $K_{\infty}$ as a function of the distance $L$, we select a pair $(\mu,\nu)$ that roughly maximizes the achievable distance and use that pair for all values of $L$. Notably, since we are considering the asymptotic regime, $K_{\infty}$ does not depend on the probabilities of the decoy settings, $p_{\nu}$ and $p_{\omega}$, nor on the probability of selecting the $X$ basis, $q_{X}$, in such a way that setting $p_{\mu} \approx 1$ and $q_{Z} \approx 1$ maximizes $K_{\infty}$.

In Fig.~\ref{fig:fixed} we illustrate the effect of the intensity correlations in a rate-distance representation for various values of the maximum relative deviation
$\delta_{\text{max}}\in\{10^{-2},10^{-3},10^{-4}\}$, and for $\xi=1$ ({\it i.e.}, for the case of nearest-neighbour pulse correlations) and $\xi=2$ (corresponding to pulse correlations of range two). The simulations show that the secret key rate is very sensitive to the deviation $\delta_{\text{max}}$. What is more, with realistic experimental data \cite{japoneses}, we find that the maximum distance attainable is less than 50 km. Moreover, it is important to remark that, as expected, the fine-grained analysis presented here outperforms the coarse-grained analysis introduced in \cite{Zapatero} regardless the value of $\delta_{\text{max}}$. Indeed, Fig.~\ref{fig:fixed} shows that when $\delta_{\text{max}}=10^{-2}$ now the attainable distance is approximately double than that in~\cite{Zapatero}. Also, note that this figure assumes a single worst-case deviation parameter $\delta_{\text{max}}$ (following Eq.~\eqref{cota_texto}). The case corresponding to Eq.~(\ref{cotavariable_texto}) is illustrated in the next section, when we compare this model with the one where Alice and Bob know the probability density function of the correlations.

\section{Truncated normal model}\label{Truncated}

\subsection{Characterization}\label{CharacterizationII}

In this section we show that a significant improvement on the secret key rate presented in Fig.~\ref{fig:fixed} can be obtained when Alice and Bob know the probability density function given by Eq.~\eqref{pns}. By assuming this, the analysis that we present now is similar to that of the previous section but simpler, as we do not have to bound the photon-number statistics, but we can solve them numerically.

The main motivation to consider this scenario is that recent decoy-state QKD experiments~\cite{japoneses} performed in the high-speed regime seem to indicate that the correlation function is not arbitrary but gaussian-shaped (see also~\cite{maka}). Triggered by this observation, for concreteness and illustration purposes we shall we assume here a truncated gaussian (TG) distribution for the correlations, which follows from renormalizing a gaussian distribution to a fixed finite interval $[\lambda,\Lambda]$. We note, however, that the analysis presented in this section could be straightforwardly adapted to any other probability density function of the correlations. To be precise, given both the mean and the variance ---say, $\gamma$ and $\sigma^{2}$ respectively---of the parental gaussian distribution together with the truncation interval $[\lambda,\Lambda]$, the TG model reads
\begin{equation}
\label{correlation_gauss}
g_{\vec{a_{k}}}(\gamma, \sigma, \lambda, \Lambda ; \alpha_{k})= \begin{cases}0 & \text { if } \alpha_{k} \leq \lambda \\ \frac{\phi\left(\gamma, \sigma^{2}; \alpha_{k}\right)}{\Phi\left(\gamma, \sigma^{2} ; \Lambda\right)-\Phi\left(\gamma,\sigma^{2} ;\lambda \right)} & \text { if } \lambda<\alpha_{k}<\Lambda \\ 0 & \text { if } \Lambda \leq \alpha_{k}\end{cases}
\end{equation}
where 
\begin{equation}
\begin{aligned}
\label{aux}
&\phi\left(\gamma, \sigma^{2} ; x\right)=\frac{1}{\sigma \sqrt{2 \pi}} e^{-\frac{(x-\gamma)^{2}}{2 \sigma^{2}}},\\
&\Phi\left(\gamma, \sigma^{2} ; x\right)=\int_{-\infty}^{x} \frac{1}{\sigma \sqrt{2 \pi}} e^{-\frac{(t-\gamma)^{2}}{2 \sigma^{2}}} d t.
\end{aligned}
\end{equation}

From now on, we shall denote by $\tilde{\gamma}$ and $\tilde{\sigma}^2$ the mean and variance of the TG distribution.
We also call $\gamma_{\vec{a_{k}}}$, the average intensity given the sequence $\vec{a_{k}}$, and $\sigma_{\vec{a_{k}}}$, the standard deviation given the sequence $\vec{a_{k}}$ of the parental normal distribution, although below the subscript indicating the record of settings will be omitted unless it can lead to confusion. More details on the truncation model are given in Appendix~\ref{appendix:truncated}.

Importantly, under suitable truncation conditions, this model reproduces the observed gaussian-shaped correlations in a finite support $[\lambda,\Lambda]$ (in contrast to the unbounded support of non-truncated gaussian distributions). We note as well that the relevant parameters can be estimated experimentally by monitoring the intensities in long sequences of rounds.

From Eq.~\eqref{correlation_gauss}, the photon-number statistics now read
\begin{equation}\label{pnst}
\left. p_{n_{k}}\right|_{\vec{a}_{k}}=\int_{\lambda}^{\Lambda}  \frac{\phi\left(\gamma_{\vec{a_{k}}}, \sigma_{\vec{a_{k}}}^{2} ; \alpha_{k}\right)e^{-\alpha_{k}} \alpha_{k}^{n_{k}}}
{\left[\Phi\left(\gamma_{\vec{a_{k}}}, \sigma_{\vec{a_{k}}}^{2} ; \Lambda\right)-\Phi\left(\gamma_{\vec{a_{k}}},\sigma_{\vec{a_{k}}}^{2} ;\lambda \right)\right]n_{k} !} \frac{}{} d \alpha_{k}.
\end{equation}
 
In general, as has been experimentally observed in~\cite{Fadri}, $\gamma_{\vec{a_{k}}}\neq{}a_{k}$ for $a_k\in\{\mu, \nu, \omega\}$, {\it i.e.} a certain displacement in the mean intensity typically occurs. We account for this shift in the signal and the decoy intensity settings, while we neglect it for the vacuum intensity setting following the observations of \cite{japoneses}. Notably as well, $\sigma_{\vec{a_{k}}}^{2}$ generally depends on the previous intensity settings, and so far no assumption about the range of the correlations has been made. In this sense, despite the fact that the analysis is valid for an arbitrary range, for simplicity here we only consider nearest-neighbour correlations for illustration purposes, {\it i.e.}, below we assume $\xi=1$.
 
Following the calculations in Appendix \ref{sec:methodsB}, and combining the analysis above with the TG model here presented, the bound on the squared overlap in the nearest-neighbour setting simply reads
\begin{equation}
\begin{aligned}
&\sum_{a_{k+1}\in A} p_{a_{k+1}}\sum_{n_{i}=0}^{\infty}\sqrt{p_{n_{i}}|_{a_{k+1},a}p_{n_{i}}|_{a_{k+1},b}}\geq\\
&\sum_{a_{k+1}\in A} p_{a_{k+1}}\sum_{n_{i}=0}^{n_{\text{cut}}}\sqrt{p_{n_{i}}|_{a_{k+1},a}p_{n_{i}}|_{a_{k+1},b}}\equiv
\sqrt{\tau_{a,b,c,n}^{\xi=1}},
\end{aligned}
\end{equation}
where we have introduced a threshold photon-number $n_{\text{cut}}$ for the numerics.

\subsection{Linear programs for parameter estimation}

The techniques for calculating the relevant parameters to evaluate the secret key rate formula using the decoy-state constraints in this model are common with those deployed in the model-independent case evaluated in the previous section. The only difference is that certain steps such as the one that leads to Eq.~\eqref{acotando} are no longer necessary. Here we solve the photon-number statistics given by Eq.~(\ref{pnst}) numerically and there is no need to invoke monotonicity arguments to bound them. The resulting linear program for the relevant single-photon yield (error click probability in the $X$ basis) reads:
\begin{widetext}
\begin{gather}\label{LPgauss_yield}
\begin{aligned}
\textup{min}\hspace{.3cm}&q_{\mathrm{Z}}^{2} p_{\mu}\sum_{h\in A}p_{h}p^{(k)}(1|\mu, h) y_{1, \mu, h, N}\\
\textup{s.t.}\hspace{.3cm}&\frac{\left\langle Z_{a,c}^{(k)}\right\rangle }{q_{\mathrm{Z}}^{2} p_{a} p_{c}}\geq\sum_{n=0}^{n_{cut}} p_{n|a,c} y_{n, a, c, N},\\
&\frac{\left\langle Z_{a,c}^{(k)}\right\rangle }{q_{\mathrm{Z}}^{2} p_{a} p_{c}} \leq 1- \sum_{n=0}^{n_{cut}} p_{n|a,c} + \sum_{n=0}^{n_{cut}} p_{n|a,c} y_{n, a, c, N},\\
&c_{a b c, n}^{+}+m_{a b c, n}^{+} y_{n, a, c, N} \geq y_{n, b, c, N}\left(a, b, c \in A, b \neq a, n=0, \ldots, n_{\mathrm{cut}}\right), \\
&c_{a b c, n}^{-}+m_{a b c, n}^{-} y_{n, a, c, N} \leq y_{n, b, c, N}\left(a, b, c \in A, b \neq a, n=0, \ldots, n_{\mathrm{cut}}\right), \\
&0 \leq y_{n, a, b, N} \leq 1\left(a, b \in A, n=0, \ldots, n_{\mathrm{cut}}\right).
\end{aligned}
\end{gather}
and
\begin{gather}\label{LPgauss_error}
\begin{aligned}
\textup{max}\hspace{.3cm}&q_{\mathrm{X}}^{2} p_{\mu}\sum_{h\in A}p_{h}p^{(k)}(1|\mu, h) h_{1, \mu, h, N} \\
\textup{s.t.}\hspace{.3cm}&\frac{\left\langle E_{a,c}^{(k)}\right\rangle }{q_{\mathrm{X}}^{2} p_{a} p_{c}}\geq\sum_{n=0}^{n_{cut}} p_{n|a,c} h_{n, a, c, N},\\
&\frac{\left\langle E_{a,c}^{(k)}\right\rangle }{q_{\mathrm{X}}^{2} p_{a} p_{c}} \leq 1- \sum_{n=0}^{n_{cut}} p_{n|a,c} + \sum_{n=0}^{n_{cut}} p_{n|a,c} h_{n, a, c, N},\\
&t_{a b c, n}^{+}+s_{a b c, n}^{+} h_{n, a, c, N} \geq h_{n, b, c, N}\left(a, b, c \in A, b \neq a, n=0, \ldots, n_{\mathrm{cut}}\right), \\
&t_{a b c, n}^{-}+s_{a b c, n}^{-} h_{n, a, c, N} \leq h_{n, b, c, N}\left(a, b, c \in A, b \neq a, n=0, \ldots, n_{\mathrm{cut}}\right), \\
&0 \leq h_{n, a, b, N} \leq 1\left(a, b \in A, n=0, \ldots, n_{\mathrm{cut}}\right).
\end{aligned}
\end{gather}
\end{widetext}

\subsection{Simulations}  

The TG model requires to experimentally determine all of its parameters, namely, the truncation ranges, the mean intensities and the standard deviations of the distribution for every combination of pulses in rounds $k$ and $k-1$ (if one assumes, as already mentioned, nearest-neighbour intensity correlations), which in turn should be experimentally accessible by monitoring the output of the transmitter in long sequences of rounds. 

As discussed in Appendix \ref{appendix:truncated}, regardless of the value of the mean intensity and the standard deviation, that in principle depend on the present and the previous settings, for illustration purposes we determine the truncation range for these simulations as $(\lambda,\Lambda)=(\tilde{\gamma}-t\tilde{\sigma},\tilde{\gamma}+t\tilde{\sigma})$, where $\tilde{\gamma}$ is the measured mean intensity of the TG distribution and $\tilde{\sigma}$ is the measured standard deviation of the TG distribution. Note that ideally, the truncation range could be measured experimentally. We can now make a direct comparison between this model in which we assume that Alice and Bob know the probability density function of the correlations, and the one that uses a maximum relative deviation defined in Eq.~\eqref{cotavariable_texto}. 

For this, we use Eq.~\eqref{skr} and  the channel model presented in Appendix~\ref{appendix:values} to evaluate the performance. Moreover, regarding the mean and the standard deviations as a function of the intensity settings in rounds $k$ and $k-1$, we take the experimental values reported in \cite{japoneses}. The only exception is the normalized standard deviation $\hat{\sigma}=\tilde{\sigma}/\tilde{\gamma}$ corresponding to vacuum, where \cite{japoneses} does not report any value and we select for illustration purposes $10^{-5}$ for all possible intensities in the round $k-1$. This is so because the fluctuation of the vacuum setting seems to be essentially negligible. In addition, for simplicity, we do not optimize the intensity settings, but fix them to $\mu=0.5$, $\nu=0.2$ and $\omega=10^{-4}$, which are the values considered in \cite{japoneses}, and we select the parameter $t=4$. The relevant parameters are summarized in Table~\ref{table:gauss}.
\begin{table}[h]
\centering
\begin{tabular}{ccc}
\hline
Pattern ($a_{k-1}, a_{k}$) & Average intensities ($\tilde{\gamma}$) & normalized SD ($\hat{\sigma}$) \\ \hline
$\mu,\mu$                         & 0.500              & 0.032                           \\
$\nu, \mu$                         & 0.510              & 0.032                           \\
$\omega,\mu$                      & 0.503              & 0.034                         \\
$\mu, \nu$                         & 0.210              & 0.070                           \\
$\nu, \nu$                         & 0.172              & 0.090                           \\
$\omega, \nu$                      & 0.165              & 0.091                            \\
$\mu/\nu/\omega, \omega$           & $10^{-4}$            & $10^{-5}$
\end{tabular}
\caption{Values for the average intensity $\tilde{\gamma}$ and normalized standard deviations $\hat{\sigma}=\tilde{\sigma}/\tilde{\gamma}$} reported in \cite{japoneses}. The only exception is the normalized standard deviation $\hat{\sigma}$ corresponding to vacuum, where \cite{japoneses} does not report any value and we select for illustration purposes $10^{-5}$. In our simulations we consider that they characterize a TG distribution.
\label{table:gauss}
\end{table}

To facilitate a fair comparison between this model and that of Eq. \eqref{cotavariable_texto}, we take 
\begin{equation}
\delta_{(a_{k}=i, a_{k-1}=j)} = t\hat{\sigma}_{i,j} = t\frac{\tilde{\sigma}_{i,j}}{\tilde{\gamma}_{i,j}}, 
\end{equation}
as explained in Appendix \ref{appendix:truncated}, where $i,j \in A$. Here, $\tilde{\gamma}_{i,j}$ ($\tilde{\sigma}_{i,j}$) refers to the average mean intensity (standard deviation) of round $k$ associated to the record $(a_{k-1},a_{k})=(j,i)$, while
$\hat{\sigma}_{i,j}=\tilde{\sigma}_{i,j}/\tilde{\gamma}_{i,j}$ are the normalized standard deviations of the TG model. In this way, we ensure that the physical intensity is bounded in the same interval in both cases. These parameters are obtained from Table \ref{table:gauss}.

\begin{figure}[H]
\centering
\includegraphics [width=8.6cm, height=6cm] {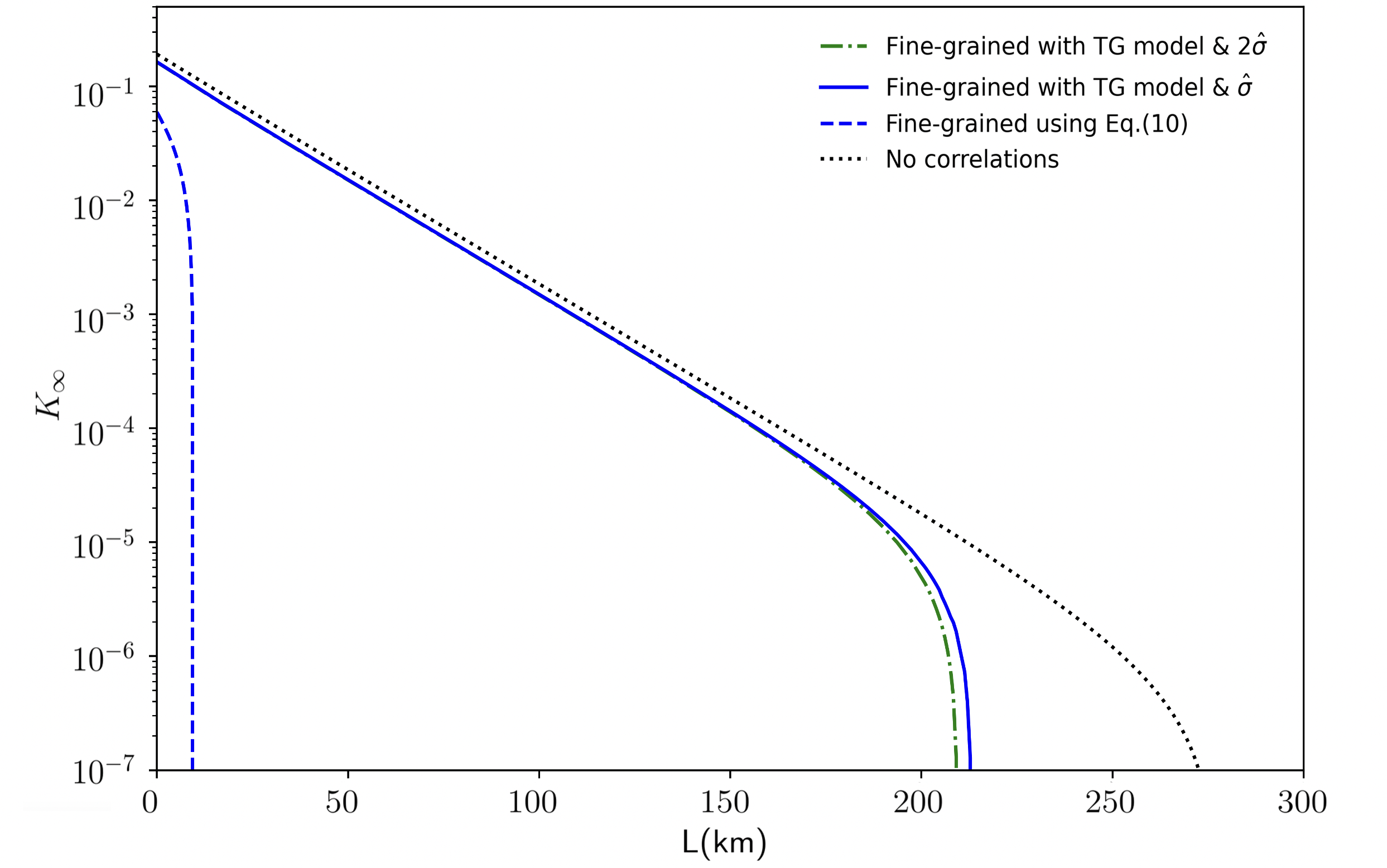}
\caption{Secret key rate within the TG model assuming the parameters from Table \ref{table:gauss} (solid blue line), and when the normalized standard deviations are twice those illustrated in that table (dash-dotted green line). For comparison, this figure also includes the model-independent case (dashed blue line) where the maximum relative deviation is taken as $\delta_{(a_{k}=i, a_{k-1}=j)}=t\hat{\sigma}_{i,j}$ for every $i,j \in A$, where $\hat{\sigma}_{i,j}$ stands for the normalized standard deviations in the TG model, and we select $t=4$. When we consider twice the normalized standard deviations provided in Table \ref{table:gauss}, this latter model provides no key.
}
\label{fig:gauss}
\end{figure}

Fig.~\ref{fig:gauss} illustrates the improvement in the secret key rate when the correlation function is known and corresponds to a TG distribution according to the model that was just introduced. We find that now the results are comparable to those of the ideal case without intensity correlations, which highlights the importance of characterizing the probability density function of the correlations. When comparing the two models studied in this paper, we conclude that knowing the distribution $g_{\vec{a_{k}}}(\alpha_{k})$ allows for a significant improvement of the resulting performance. This is due to the fact that in this scenario one can evaluate the photon-number statistics exactly, which makes the parameter estimation much tighter, improving both the decoy-state constraints and the CS constraints. Finally, Fig.~\ref{fig:gauss} also includes a representation where the standard deviations are twice the values given in Table \ref{table:gauss}, to evaluate the effect that this parameter has on the secret key rate. The fine-grained analysis described in Eq. \eqref{cotavariable_texto} does not provide key in this latter scenario. Likewise, if we use Eq.~(\ref{cota_texto}) in the model-independent case, no key is obtained in neither case.

\section{Conclusions}\label{Discussion}

When combining high-speed clock rate QKD transmitters with the decoy-state technique, the presence of intensity correlations in the generated pulses invalidate the central security assumption of this method, namely, that the yields and error rates associated to $n$-photon pulses are independent of the intensity settings. This problem can be solved by imposing constraints on the intensity-setting dependence of these parameters, which is done here by invoking the so-called Cauchy-Schwarz constraints. Essentially, these constraints arise from the indistinguishability of non-orthogonal quantum states and can be derived on the basis of a minor characterization of the correlations. 

In this work, by using a standard decoy-state BB84 protocol with three possible intensity settings, we have evaluated the effect that such correlations have in the secret key rate. We have done so by introducing a {\it fine-grained analysis} able to handle arbitrary long-range correlations. This approach leads to  significantly tighter constraints for the parameter estimation and notably improves the secret key rate achieved by previous works. Also, we have shown that the characterization of the probability density function of the correlations permits to notably improve the resulting performance, which now becomes comparable to that of the ideal scenario without intensity correlations. Putting it all together, the present work provides a solid step forward towards full implementation security in QKD with high performance.

\begin{acknowledgments}
This work was supported by Cisco Systems Inc., the Galician Regional Government (consolidation of Research Units: AtlantTIC), the Spanish Ministry of Economy and Competitiveness (MINECO), the Fondo Europeo de Desarrollo Regional (FEDER) through Grant No. PID2020-118178RB-C21, and MICIN with funding from the European Union NextGenerationEU (PRTR-C17.I1) and the Galician Regional Government with own funding through the “Planes Complementarios de I+D+I con las Comunidades Autónomas” in Quantum Communication.
\end{acknowledgments}

\appendix
\section{Cauchy-Schwarz constraint}\label{sec:methodsA}

The CS constraint can be stated as follows:
\begin{theorem}
Let $|u\rangle$ and $|v\rangle$ be pure states of a certain quantum system. Then, for all positive operators $\hat{O} \leq I$,
\begin{equation}
\label{generalCS}
G_{-}\left(\langle u|\hat{O}|u\rangle,|\langle v |u\rangle|^{2}\right) \leq\langle v|\hat{O}|v\rangle \leq G_{+}\left(\langle u|\hat{O}|u\rangle,|\langle v|u\rangle|^{2}\right),
\end{equation}
where the functions $G_{\pm}$ are given in Eq.~\eqref{eq6}.
\end{theorem}
{\bf Proof}: see Supplementary Materials of \cite{QKD_correlated}.

This constraint can be used to impose a restriction on the maximum bias that Eve can induce between the $n$-photon yields and error probabilities associated to different intensity settings.
Here we derive a linear version of Eq.~\eqref{cs_text}, enabling the use of linear programming for the decoy-state parameter estimation procedure. Following \cite{Zapatero}, we have that in virtue of the convexity/concavity of the functions that define the constraints, their first-order expansions around any given reference value provide valid linear bounds as well. Thus, if we focus on the yields, the linearization provides the following bounds:
\begin{widetext}
 \begin{eqnarray}
 G_{-}(Y_{n, v_{0}...v_{\xi}}^{(k)}, \tau_{v_{0} w_{0}...v_{\xi}, n}^{\xi}) &\geq& G_{-}(\tilde{Y}_{n, v_{0}...v_{\xi}}^{(k)}, \tau_{v_{0} w_{0}...v_{\xi}, n}^{\xi})+
 G^{\prime}_{-}(\tilde{Y}_{n, v_{0}...v_{\xi}}^{(k)}, \tau_{v_{0} w_{0}...v_{\xi}, n})(Y_{n, v_{0}...v_{\xi}}^{(k)}-\tilde{Y}_{n, v_{0}...v_{\xi}}^{(k)}), \nonumber\\
 G_{+}(Y_{n, v_{0}...v_{\xi}}^{(k)}, \tau_{v_{0} w_{0}...v_{\xi}, n}^{\xi}) &\leq& G_{+}(\tilde{Y}_{n, v_{0}...v_{\xi}}^{(k)}, \tau_{v_{0} w_{0}...v_{\xi}, n}^{\xi})+
 G^{\prime}_{+}(\tilde{Y}_{n, v_{0}...v_{\xi}}^{(k)}, \tau_{v_{0} w_{0}...v_{\xi}, n})(Y_{n, v_{0}...v_{\xi}}^{(k)}-\tilde{Y}_{n, v_{0}...v_{\xi}}^{(k)}),
\end{eqnarray}
\end{widetext}
for every $Y_{n, v_{0}...v_{\xi}}^{(k)} \in[0,1]$ independently of which reference $\tilde{Y}_{n, v_{0}...v_{\xi}}^{(k)} \in[0,1]$ is selected. 

Notice that the derivative functions $G^{\prime}_{\pm}$ are well defined for all $Y_{n, v_{0}...v_{\xi}}^{(k)} \in[0,1]$; their expression is given by Eq.~(\ref{brussels}).
Thus, given a reference yield $\tilde{Y}_{n,w_{0}...v_{\xi} }^{(k)}$, the linearized bounds read
\begin{widetext}
\begin{eqnarray}
\label{linearyield_app}
G_{-}\left(\tilde{Y}_{n,v_{0}...v_{\xi} }^{(k)}, \tau_{v_{0}w_{0}...v_{\xi}, n}^{\xi}\right)+
G_{-}^{\prime}\left(\tilde{Y}_{n, v_{0}...v_{\xi}}^{(k)}, \tau_{v_{0}w_{0}...v_{\xi}, n}^{\xi}\right)\left(Y_{n, v_{0}...v_{\xi}}^{(k)}-\tilde{Y}_{n, v_{0}...v_{\xi}}^{(k)}\right)
&\leq& Y_{n, w_{0}...v_{\xi}}^{(k)}, \nonumber\\
G_{+}\left(\tilde{Y}_{n, v_{0}...v_{\xi}}^{(k)}, \tau_{v_{0}w_{0}...v_{\xi}, n}^{\xi}\right)+
G_{+}^{\prime}\left(\tilde{Y}_{n, v_{0}...v_{\xi}}^{(k)}, \tau_{a w_{0}...v_{\xi}, n}^{\xi}\right)\left(Y_{n, v_{0}...v_{\xi}}^{(k)}-\tilde{Y}_{n, v_{0}...v_{\xi}}^{(k)}\right)
&\geq&Y_{n, w_{0}...v_{\xi}}^{(k)}.
\end{eqnarray}
\end{widetext}

Similarly for the error probabilities, and assuming that we select reference parameters independent of Alice's bit value $r_{k}$ (for a more detailed analysis see \cite{Zapatero}) we have:
\begin{widetext}
\begin{eqnarray}
\label{linearerror_app}
G_{-}\left(\tilde{H}_{n,v_{0}...v_{\xi} }^{(k)}, \tau_{v_{0}w_{0}...v_{\xi}, n}^{\xi}\right)+
G_{-}^{\prime}\left(\tilde{H}_{n, v_{0}...v_{\xi}}^{(k)}, \tau_{v_{0}w_{0}...v_{\xi}, n}^{\xi}\right)\left(H_{n, v_{0}...v_{\xi}}^{(k)}-\tilde{H}_{n, v_{0}...v_{\xi}}^{(k)}\right)
&\leq& H_{n, w_{0}...v_{\xi}}^{(k)}, \nonumber\\
G_{+}\left(\tilde{H}_{n, v_{0}...v_{\xi}}^{(k)}, \tau_{v_{0}w_{0}...v_{\xi}, n}^{\xi}\right)+
G_{+}^{\prime}\left(\tilde{H}_{n, v_{0}...v_{\xi}}^{(k)}, \tau_{a w_{0}...v_{\xi}, n}^{\xi}\right)\left(H_{n, v_{0}...v_{\xi}}^{(k)}-\tilde{H}_{n, v_{0}...v_{\xi}}^{(k)}\right)
&\geq& H_{n, w_{0}...v_{\xi}}^{(k)}.
\end{eqnarray}
\end{widetext}

To finish with, note that one can restrict the reference parameters to be round-independent, in such a way that, summing over $k$ and dividing by $N$ in Eqs.~\eqref{linearyield} and~\eqref{linearerror}, we obtain the average parameters 
\begin{eqnarray}
y_{n, w_{0}...v_{\xi}, N}&=&\sum_{k=1}^{N} \frac{Y_{n, w_{0}...v_{\xi}}^{(k)}}{N}, \nonumber \\
h_{n, w_{0}...v_{\xi}, N}&=&\sum_{k=1}^{N} \frac{H_{n, w_{0}...v_{\xi}}^{(k)}}{N}, 
\end{eqnarray}
and similarly for the other terms that appear below. 

We define the intercepts and slopes as:
\begin{eqnarray}
\label{esesta}
c_{v_{0}w_{0}...v_{\xi}, n}^{\pm}&=&G_{\pm}\left(\tilde{y}_{n, v_{0}...v_{\xi}}, \tau_{v_{0}w_{0}...v_{\xi}, n}^{\xi}\right)\nonumber\\
&-&G_{\pm}^{\prime}\left(\tilde{y}_{n, v_{0}...v_{\xi}}, \tau_{v_{0}w_{0}...v_{\xi}, n}^{\xi}\right) \tilde{y}_{n, v_{0}...v_{\xi}}, \nonumber \\ 
m_{v_{0}w_{0}...v_{\xi}, n}^{\pm}&=&G_{\pm}^{\prime}\left(\tilde{y}_{n, v_{0}...v_{\xi}}, \tau_{v_{0}w_{0}...v_{\xi}, n}^{\xi}\right),\nonumber \\
t_{v_{0}w_{0}...v_{\xi}, n}^{\pm}&=&G_{\pm}\left(\tilde{h}_{n, v_{0}...v_{\xi}}, \tau_{v_{0}w_{0}...v_{\xi}, n}^{\xi}\right)\nonumber\\
&-&G_{\pm}^{\prime}\left(\tilde{h}_{n, v_{0}...v_{\xi}}, \tau_{v_{0}w_{0}...v_{\xi}, n}^{\xi}\right) \tilde{h}_{n, v_{0}...v_{\xi}}\nonumber \\ 
s_{v_{0}w_{0}...v_{\xi}, n}^{\pm}&=&G_{\pm}^{\prime}\left(\tilde{h}_{n, v_{0}...v_{\xi}}, \tau_{v_{0}w_{0}...v_{\xi}, n}^{\xi}\right).
\end{eqnarray}

The linear CS constraints for the round independent case are:
\begin{widetext}
\begin{eqnarray}
c_{v_{0}w_{0}...v_{\xi}, n}^{+}+m_{v_{0}w_{0}...v_{\xi}, n}^{+} y_{n, v_{0}... v_{\xi}, N} &\geq& y_{n, w_{0}... v_{\xi}, N}\left(v_{0},w_{0},...,v_{\xi} \in A, w_{0} \neq v_{0}\right),\nonumber \\
c_{v_{0}w_{0}...v_{\xi}, n}^{-}+m_{v_{0}w_{0}...v_{\xi}, n}^{-} y_{n, v_{0}... v_{\xi}, N} &\leq& y_{n, w_{0}... v_{\xi}, N}\left(v_{0},w_{0},...,v_{\xi} \in A, w_{0} \neq v_{0}\right),\nonumber \\
0 &\leq& y_{n, v_{0}... v_{\xi}, N} \leq 1\left(v_{0,}....,v_{\xi} \in A\right),
\end{eqnarray}
and
\begin{eqnarray}
t_{v_{0}w_{0}...v_{\xi}, n}^{+}+s_{v_{0}w_{0}...v_{\xi}, n}^{+} h_{n, v_{0}... v_{\xi}, N} &\geq& h_{n, w_{0}... v_{\xi}, N}\left(v_{0},w_{0},...,v_{\xi} \in A, w_{0} \neq v_{0}\right),\nonumber \\
t_{v_{0}w_{0}...v_{\xi}, n}^{-}+s_{v_{0}w_{0}...v_{\xi}, n}^{-} h_{n, v_{0}... v_{\xi}, N} &\leq& h_{n, w_{0}... v_{\xi}, N}\left(v_{0},w_{0},...,v_{\xi} \in A, w_{0} \neq v_{0}\right),\nonumber \\
0 &\leq& h_{n, v_{0}... v_{\xi}, N} \leq 1\left(v_{0,}....,v_{\xi} \in A\right).
\end{eqnarray}
\end{widetext}

Note that the characterization of the quantum channel is essential, as  the tightness of these linear bounds is subject to the adequacy of the selected reference parameters, and so it relies on it. The selected reference values are presented in Appendix \ref{appendix:values}.

\section{Derivation of $\tau_{v_{0},w_{0},...v_{\xi}, n}^{\xi}$}\label{Methods}

In an entanglement based view of the protocol, the global input state describing all the protocol rounds reads:
\begin{eqnarray}
\label{estado}
|\Psi\rangle&=&\biggl[\sum_{a_{1}^{N}} \sum_{x_{1}^{N}} \sum_{r_{1}^{N}}\left(\prod_{i=1}^{N} \sqrt{\frac{p_{a_{i}} q_{x_{i}}}{2}}\right)\biggl(\bigotimes_{i=1}^{N}\left|a_{i} \right\rangle_{A_{i}}\left|x_{i}\right\rangle_{\tilde{A}_{i}}\nonumber\\
&\otimes&\left|r_{i}\right\rangle_{A_{i}^{\prime}}\left|\psi_{\vec{a}_{i}}^{x_{i}, r_{i}}\right\rangle_{B_{i} C_{i}}\biggl)\biggl] \otimes|0\rangle_{E},
\end{eqnarray}
where $a_{1}^{N}=a_{1} \ldots a_{N}$ and equivalently for $x_{1}^{N}$ and $r_{1}^{N}$. Also, for every round $i$, $\{\left|a_{i}\right\rangle_{A_{i}}|a_{i} \in A\}$, $\{\left|x_{i}\right\rangle_{\tilde{A}_{i}}|x_{i} \in B\}$ and $\{\left|r_{i}\right\rangle_{A_{i}^{\prime}}| r_{i} \in \mathbb{Z}_{2}\}$ are orthonormal bases of Alice’s $i$-th registers $A_{i}$, $\tilde{A}_{i}$ and $A_{i}^{\prime}$. The state $|0\rangle_{E}$ stands for the initial state of Eve's ancillary system. We also define:
\begin{equation}
\left|\psi_{\vec{a}_{i}}^{x_{i}, r_{i}}\right\rangle_{B_{i} C_{i}}=\sum_{n_{i}=0}^{\infty} \sqrt{p_{n_{i}\mid \vec{a}_{i}}}\left|t_{n_{i}}\right\rangle_{C_{i}}\left|n_{i}^{x_{i}, r_{i}}\right\rangle_{B_{i}},
\end{equation}
where $C_{i}$ denotes an inaccessible purifying system with orthonormal basis $\{\left|t_{n_{i}}\right\rangle_{C_{i}} |n_{i} \in \mathbb{N}\}$ ($C_{i}$ stores the photon-number information of the $i$-th signal that Alice sends to Bob), $B_{i}$ denotes the system delivered to Bob, $|n_{i}^{x_{i}, r_{i}}\rangle_{B_{i}}$ stands for a Fock state with $n_{i}$ photons encoding the BB84 polarization state defined by ($x_{i}, r_{i}$), and the photon-number statistics $\left.p_{n_{i}}\right|_{\vec{a}_{i}}$ are defined in Eq.~\eqref{pns}.

Let us denote by $\hat{U}_{B E}$ Eve's coherent interaction with systems $B_{1},...,B_{N}$
and $E$ so that $\hat{U}_{B E}|\Psi\rangle$ represents the global state prior to Bob’s measurements. We also refer to Bob's click POVM element in round $k$ as 
\begin{equation}
\hat{M}^{\text{click}}_{B_{k}}=\mathbb{I}_{B_{k}}-\hat{M}_{B_{k}}^{f}. 
\end{equation}
The joint probability $p^{(k)}(\text {click}, n, v_{0},...,v_{\xi} , \mathrm{Z})$ is computed as:
\begin{widetext}
\begin{eqnarray}
\label{joint}
p^{(k)}(\text {click}, n, v_{0},...,v_{\xi} , Z)&=&\operatorname{Tr}\left\{\hat{P}_{\left|v_{0},..., v_{\xi}, Z, t_{n}\right\rangle_{A_{k} ... A_{k-\xi} \tilde{A}_{k} C_{k}}} \hat{M}_{B_{k}}^{\text {click }} \hat{U}_{B E}|\Psi\rangle\langle\Psi| \hat{U}_{B E}^{\dagger}\right\}\nonumber \\
&=&\operatorname{Tr}\left\{\hat{U}_{B E}^{\dagger} \hat{M}_{B_{k}}^{\text {click }} \hat{U}_{B E} \hat{P}_{\left|v_{0},..., v_{\xi}, Z, t_{n}\right\rangle_{A_{k}... A_{k-\xi} \tilde{A}_{k} C_{k}}}|\Psi\rangle\langle\Psi| \hat{P}_{\left|v_{0},..., v_{\xi}, Z, t_{n}\right\rangle_{A_{k}... A_{k-\xi} \tilde{A}_{k} C_{k}}}\right\}\nonumber \\
&=&\operatorname{Tr}_{\underline{A_{k} \tilde{A}_{k} C_{k}}, A_{1}^{\prime N} B_{1}^{N} E}\left\{\hat{U}_{B E}^{\dagger} \hat{M}_{B_{k}}^{\text {click }} \hat{U}_{B E}\left|\widetilde{\Psi}_{v_{0},...,v_{\xi} Z, n}^{(k,...,k-\xi)}\right\rangle\left\langle\widetilde{\Psi}_{v_{0},...,v_{\xi}, Z, n}^{(k,...,k-\xi)}\right|\right\},
\end{eqnarray}
\end{widetext}
where 
\begin{widetext}
\begin{eqnarray}
\hat{P}_{\left|v_{0},...,v_{\xi}, Z, t_{n}\right\rangle_{A_{k} \tilde{A}_{k}... A_{k-\xi} C_{k}}}&=&|v_{\xi}\rangle\langle v_{\xi} |_{A_{k-\xi}}\otimes...\otimes |v_{0}\rangle\langle v_{0}|_{A_{k}}\otimes |Z\rangle\langle Z|_{\tilde{A}_{k}} \otimes | t_{n}\rangle\langle\left. t_{n}\right|_{C_{k}}, 
\end{eqnarray}
\end{widetext}
$\underline{A_{k}}=\{A_{j} | j \neq k,.., k-\xi\}$, $\underline{\tilde{A}_{k}}=\{\tilde{A}_{j} |j \neq k\}$ and $\underline{C_{k}}=\left\{C_{j} |j \neq k \right\} $. Note that, for round $k$, we project on the intensity setting, the basis and the photon-number, and for all previous rounds, we only project on the intensity setting. The unnormalized pure state is defined as:
\begin{equation}
\label{proy}
\left|\widetilde{\Psi}_{v_{0},...,v_{\xi}, Z, n}^{(k,...,k-\xi)}\right\rangle=\langle v_{\xi}|_{A_{k-\xi}}...\langle v_{0}|_{A_{k}} \langle Z|_{\tilde{A}_{k}}  \langle t_{n}|_{C_{k}} |\Psi\rangle.
\end{equation}

We now re-state  Eq.~\eqref{joint} in terms of the normalized state 
\begin{equation}
\left|\Psi_{v_{0},...,v_{\xi}, \mathrm{Z}, n}^{(k,...,k-\xi)}\right\rangle=\frac{\left|\widetilde{\Psi}_{v_{0},...,v_{\xi}, \mathrm{Z}, n}^{(k,...,k-\xi)}\right\rangle}{\left\|\left|\widetilde{\Psi}_{v_{0},...,v_{\xi}, \mathrm{Z}, n}^{(k,...,k-\xi)}\right\rangle \right\|}, 
\end{equation}
which leads to:
\begin{widetext}
\begin{equation}
p^{(k)}(\text {click, } n, v_{0},...,v_{\xi}, \mathrm{Z})=\left\|\left|\widetilde{\Psi}_{v_{0},...,v_{\xi}, \mathrm{Z}, n}^{(k,...,k-\xi)}\right\rangle \right\|^{2}\operatorname{Tr}\left\{\hat{U}_{B E}^{\dagger} \hat{M}_{B_{k}}^{\text {click }} \hat{U}_{B E}\left|\Psi_{v_{0},...v_{\xi}, \mathrm{Z}, n}^{(k,...,k-\xi)}\right\rangle\left\langle\Psi_{v_{0},...v_{\xi}, \mathrm{Z}, n}^{(k,...,k-\xi)}\right|\right\}.
\end{equation}
\end{widetext}

We note that $p^{(k)}(n, v_{0},...,v_{\xi}, Z)$ is given by
$p^{(k)}(n, v_{0},...,v_{\xi}, Z)=\operatorname{Tr}\{\hat{P}_{\left|v_{0},...,v_{\xi}, Z, t_{n}\right\rangle_{A_{k} C_{k}}} \hat{U}_{B E}|\Psi\rangle\langle\Psi| \hat{U}_{B E}^{\dagger}\}=\left\|\left|\widetilde{\Psi}_{v_{0},...,v_{\xi}, Z, n}^{(k,...,k-\xi)}\right\rangle\right\|^{2}$. Therefore, in virtue of Bayes rule we obtain:
\begin{eqnarray}
p^{(k)}(\text {click}|n, v_{0},...,v_{\xi}, \mathrm{Z})&=&\big\langle\Psi_{v_{0},...,v_{\xi}, \mathrm{Z}, n}^{(k,...,k-\xi)}\big|\hat{O}_{\text {click }}^{(k)}\nonumber \\
&\times&\big| \Psi_{v_{0},...,v_{\xi}, \mathrm{Z}, n}^{(k,...,k-\xi)}\big\rangle,
\end{eqnarray}
where we have defined 
\begin{equation}
\hat{O}_{\text {click }}^{(k)}=\hat{U}_{B E}^{\dagger} \hat{M}_{B_{k}}^{\text {click }} \hat{U}_{B E}. 
\end{equation}

Now, from Eq.~\eqref{yield}, we recall that the quantity defined above is actually the $n$-photon yield of round $k$ associated to the record of settings $v_{0},...,v_{\xi}$. From the CS constraint, it follows that:
\begin{widetext}
\begin{equation}
G_{-}\left(Y_{n, v_{0}...v_{\xi}}^{(k)},\left|\left\langle\Psi_{w_{0}...v_{\xi}, \mathrm{Z}, n}^{(k,...,k-\xi)} |\Psi_{v_{0}...v_{\xi}, \mathrm{Z}, n}^{(k,...,k-\xi)}\right\rangle\right|^{2}\right) \leq Y_{n, w_{0}...v_{\xi}}^{(k)} \leq 
G_{+}\left(Y_{n, v_{0}...v_{\xi}}^{(k)},\left|\left\langle\Psi_{w_{0}...v_{\xi}, \mathrm{Z}, n}^{(k,...,k-\xi)}| \Psi_{v_{0}...v_{\xi}, \mathrm{Z}, n}^{(k,...,k-\xi)}\right\rangle\right|^{2}\right),
\end{equation}
\end{widetext}
for all $n \in \mathbb{N}$, $w_{0},v_{0},...,v_{\xi}\in A$ where $v_{0}\neq w_{0}$ and $k=1,...,N$. Note that the closer the inner product between the states is to one, the tighter the bounds of the inequality are.

Having reached this stage, we recall that the goal is to set a lower bound on the inner product $\langle\Psi_{w_{0}...v_{\xi}, \mathrm{Z}, n}^{(k,...,k-\xi)} \mid \Psi_{v_{0}...v_{\xi}, \mathrm{Z}, n}^{(k,...,k-\xi)}\rangle$. From Eq.~\eqref{estado} and Eq.~\eqref{proy}, direct calculation shows that:
\begin{widetext}
\begin{eqnarray}
\left|\widetilde{\Psi}_{v_{0},v_{1}...v_{\xi}, \mathrm{Z}, n}^{(k,...,k-\xi)}\right\rangle&=&\sqrt{\frac{q_{\mathrm{z}} p_{v_{0}}... p_{v_{\xi}}}{2^{N}}}\Bigg[\sum_{\underline{a_{k}}} 
\sum_{\underline{x_{k}}} \sum_{r_{1}^{N}}\left(\prod_{i \neq k} \sqrt{q_{x_{i}}}\right)\left(\prod_{i \neq k... k-\xi} \sqrt{p_{a_{i}}}\right)
\bigotimes_{i=1}^{k-\xi-1}\left|a_{i} \right\rangle_{A_{i}}\left|x_{i}\right\rangle_{\tilde{A}_{i}}\left|r_{i}\right\rangle_{A_{i}^{\prime}}\left|\psi_{\vec{a}_{i}}^{x_{i}, r_{i}}\right\rangle_{B_{i} C_{i}} \nonumber\\
&\otimes&\left|x_{k-\xi}\right\rangle_{\tilde{A}_{k-\xi}}\left|r_{k-\xi}\right\rangle_{A_{k-\xi}^{\prime}}\left|\psi_{\vec{a}_{k-\xi}(a_{k-\xi}=v_{\xi})}^{x_{k-\xi}, r_{k-\xi}}\right\rangle_{B_{k-\xi} C_{k-\xi}} \left|x_{k-\xi+1}\right\rangle_{\tilde{A}_{k-\xi+1}}\left|r_{k-\xi+1}\right\rangle_{A_{k-\xi+1}^{\prime}}\nonumber\\ 
&\otimes&\left|\psi_{\vec{a}_{k-\xi+1}(a_{k-\xi+1}=v_{\xi-1},a_{k-\xi}=v_{\xi})}^{x_{k-\xi+1}, r_{k-\xi+1}}\right\rangle_{B_{k-\xi+1} C_{k-\xi+1}} \ldots\left|x_{k-1}\right\rangle_{\tilde{A}_{k-1}}\left|r_{k-1}\right\rangle_{A_{k-1}^{\prime}} \nonumber\\
&\otimes&\left|\psi_{\vec{a}_{k-1}(a_{k-1}=v_{1}...a_{k-\xi}=v_{\xi})}^{x_{k-1}, r_{k-1}}\right\rangle_{B_{k-1} C_{k-1}}\sqrt{\left.p_{n}\right|_{v_{0},v_{1},...,v_{\xi},\vec{a}_{k-\xi-1}}}\left|r_{k}\right\rangle_{A_{k}^{\prime}}\left|n^{\mathrm{Z}, r_{k}}\right\rangle_{B_{k}}
\bigotimes_{i=k+1}^{N}\left|a_{i} \right\rangle_{A_{i}}\left|x_{i} \right\rangle_{\tilde{A}_{i}}\left|r_{i}\right\rangle_{A_{i}^{\prime}}\nonumber \\
&\otimes&\left|\psi_{\vec{a}_{i}\left(a_{k}=v_{0},..., a_{k-\xi}=v_{\xi}\right)}^{x_{i}, r_{i}}\right\rangle_{B_{i} C_{i}}\Bigg] \otimes|0\rangle_{E},
\end{eqnarray}
\end{widetext}
where $\underline{x_{k}}=\left\{x_{j}|j \neq k, \right\}$ and $\underline{a_{k}}=\left\{a_{j}| j \neq k,..., k-\xi\right\}$. Then, computing the inner product yields:
\begin{widetext}
\begin{equation}
\label{inter1}
\begin{aligned}
&\left\langle\widetilde{\Psi}_{w_{0},v_{1},..., v_{\xi}, Z, n}^{(k,...,k-\xi)}|\widetilde{\Psi}_{v_{0},v_{1},..., v_{\xi}, Z, n}^{(k,...,k-\xi)}\right\rangle=\frac{q_{z} \sqrt{p_{v_{0}}...p_{v_{\xi}}p_{w_{0}}...p_{v_{\xi}}}}{2^{N}} \sum_{\underline{a_{k}}} \sum_{\underline{x_{k}}} \sum_{r_{1}^{N}}\left(\prod_{i \neq k} q_{x_{i}}\right) \left(\prod_{i \neq k,..,k-\xi} p_{a_{i}}\right) \\
&\left\langle \psi_{\vec{a}_{k-\xi}\left(a_{k-\xi}=v_{\xi}\right)}^{x_{k-\xi}, r_{k-\xi}} | \psi_{\vec{a}_{k-\xi}\left(a_{k-\xi}=v_{\xi}\right)}^{x_{k-\xi}, r_{k-\xi}}\right\rangle_{B,C_{k-\xi} }\left\langle \psi_{\vec{a}_{k-\xi+1}\left(a_{k-\xi+1}=v_{\xi-1}, a_{k-\xi}=v_{\xi}\right)}^{x_{k-\xi+1}, r_{k-\xi+1}}| \psi_{\vec{a}_{k-\xi+1}\left(a_{k-\xi+1}=v_{\xi-1}, a_{k-\xi+1}=v_{\xi-1}\right)}^{x_{k-\xi+1}, r_{k-\xi+1}}\right\rangle_{B,C_{k-\xi+1}} \\
&\ldots\left(\sqrt{p_{n}|_{v_{0},v_{1},..., v_{\xi}} p_{n}|_{w_{0},v_{1},..., v_{\xi}}}\right)\left(\prod_{i=k+1}^{\min\{k+\xi,N\}} \left\langle \psi_{\vec{a}_{i}\left(a_{k}=w_{0},..., a_{k-\xi}=v_{\xi} \right)}^{x_{i}, r_{i}}| \psi_{\vec{a}_{i}\left(a_{k}=v_{0},..., a_{k-\xi}=v_{\xi}\right)}^{x_{i}, r_{i}}\right\rangle_{B_{i} C_{i}}\right),
\end{aligned}
\end{equation}
\end{widetext}
where we have omitted the subscript $\vec{a}_{k-\xi-1}$ in the square root term in parentheses because the photon-number statistics are independent of this sub-string of settings. Since all the factors previous to round $k$ are equal to 1, and the sums over $x_{k}$ and $r_{1}^{N}$ yield $\sum_{\underline{x_{k}}}\sum_{r_{1}^{N}}\left(\prod_{i \neq k} q_{x_{i}}\right)=\sum_{r_{1}^{N}}\left\{\sum_{\underline{x_{k}}}\left(\prod_{i \neq k} q_{x_{i}}\right)\right\}=2^{N}$, the previous equation reduces to:
\begin{widetext}
\begin{eqnarray}
\label{inter2}
\left\langle\widetilde{\Psi}_{w_{0},v_{1},..., v_{\xi}, Z, n}^{(k,...,k-\xi)}| \widetilde{\Psi}_{v_{0},v_{1},..., v_{\xi}, Z, n}^{(k,...,k-\xi)}\right\rangle&=&q_{z} \sqrt{p_{v_{0}}
p_{w_{0}}} p_{v_{1}}...p_{v_{\xi}}
\sum_{a_{\max\{1,k-2\xi\}}^{k-\xi-1}}\left(\prod_{i=\max\{1,k-2\xi\}}^{k-\xi-1} p_{a_{i}}\right) \sqrt{p_{n}|_{v_{0},..., w_{\xi}} p_{n}|_{w_{0},..., w_{\xi}}}\nonumber \\ 
&\times&\left[\sum_{a_{k+1}^{\min\{k+\xi,N\}}}\left(\prod_{i=k+1}^{\min\{k+\xi,N\}} p_{a_{i}}\left\langle\psi_{\vec{a}_{i}\left(a_{k}=w_{0},..., a_{k-\xi}=v_{\xi}\right)}^{x_{i}, r_{i}}| \psi_{\vec{a}_{i}\left(a_{k}=v_{0},...a_{k-\xi}=v_{\xi}\right)}^{x_{i}, r_{i}}\right\rangle_{B_{i} C_{i}}\right)\right], \nonumber \\
\end{eqnarray}
\end{widetext}

In terms of the normalized states, we obtain:
\begin{widetext}
\begin{eqnarray}
\label{normalization}
\left\langle\Psi_{w_{0},..., v_{\xi}, Z, n}^{(k,...,k-\xi)}| \Psi_{v_{0},..., v_{\xi}, Z, n}^{(k,...,k-\xi)}\right\rangle =\sum_{a_{k+1}^{\min\{k+\xi,N\}}}\left(\prod_{i=k+1}^{\min\{k+\xi,N\}} p_{a_{i}}\right)\left\langle\psi_{\vec{a}_{i}\left(a_{k}=w_{0},..., a_{k-\xi}=v_{\xi}\right)}^{x_{i}, r_{i}} | \psi_{\vec{a}_{i}\left(a_{k}=v_{0},...a_{k-\xi}=v_{\xi}\right)}^{x_{i}, r_{i}}\right\rangle_{B_{i} C_{i}}.\quad\quad
\end{eqnarray}
\end{widetext}

We can express Eq.~\eqref{normalization} in terms of the photon-number statistics instead:
\begin{widetext}
\begin{equation}
\label{poisson}
\left\langle\Psi_{w_{0},..., v_{\xi}, Z, n}^{(k,...,k-\xi)}| \Psi_{v_{0},..., v_{\xi},Z,n}^{(k,...,k-\xi)}\right\rangle=\sum_{a_{k+1}^{\min\{k+\xi,N\}}}\left(\prod_{i=k+1}^{\min\{k+\xi,N\}} p_{a_{i}}\right)\sum_{m=0}^{\infty}\sqrt{p_{m}|_{a_{i}...v_{0}...v_{\xi+k-i}}p_{m}|_{a_{i}...w_{0}...v_{\xi+k-i}}}.
\end{equation}
\end{widetext}
That is, this quantity takes the same value given by Eq.~(\ref{poisson}) for all $n$.

\subsection{Model-independent correlations}

Eq.~\eqref{poisson}  provides a general formula that can be used as the input of the CS constraint regardless of the correlation function. Here we consider the model-independent scenario where the correlation function is assumed to be unknown. This implies that we cannot evaluate the photon-number statistics on which Eq.~\eqref{poisson} depends. Therefore, we bound these statistics by invoking monotonicity arguments.

Only to keep the notation simple, we consider that the mean physical intensity matches the actual intensity setting, even though according to experiments, there might to be a certain displacement between these two quantities \cite{japoneses}. We emphasize, however, that such shift could be straightforwardly included in our analysis. Let us start by introducing the following shorthand notation:
\begin{equation}
\begin{aligned}
&a_{i}^{(\tau) \pm}=a_{i}\left(1 \pm \delta_{\left(a_{i}, \ldots,a_{k+1}, a_{k}=\tau,a_{k-1}=v_{1}, \ldots, a_{i-\xi}=v_{\xi+k-i}\right)}^{\pm}\right),\\
\end{aligned}
\end{equation}
with $\tau\in\{v_{0},w_{0}\}$.

From the definition of the photon-number statistics and by noticing that $e^{-x}x^{n}$ is strictly decreasing for $n=0$  and increasing for $n\geq{}1$ in $x\in(0,1)$, we have that 
\begin{eqnarray}
p_{0}|_{ \vec{a}_{i}\left(a_{k}=v_{0}\right)} &\geq& e^{-a_{i}^{(v_{0})+}},\nonumber \\
p_{n \geq 1}|_{\vec{a}_{i}\left(a_{k}=v_{0}\right)} &\geq& e^{-a_{i}^{(v_{0})-}} \frac{\left(a_{i}^{(v_{0})-}\right)^{n}}{n!}. 
\end{eqnarray}

Therefore, we can bound the square root of Eq.~\eqref{poisson} as (where for convenience we now use the variable $n$ to name the index of the sum):
\begin{widetext}
\begin{eqnarray}
\label{qwe}
\sum_{n=0}^{\infty}\sqrt{p_{n}|_{a_{i}...v_{0}...v_{\xi+k-i}}p_{n}|_{a_{i}...w_{0}...v_{\xi+k-i}}} &=& \sqrt{p_{0}|_{a_{i}...v_{0}...v_{\xi+k-i}}p_{0}|_{a_{i}...w_{0}...v_{\xi+k-i}}}+\sum_{n=1}^{\infty}\sqrt{p_{n}|_{a_{i}...v_{0}...v_{\xi+k-i}}p_{n}|_{a_{i}...w_{0}...v_{\xi+k-i}}}\nonumber \\
&\geq&
\biggl[e^{\frac{1}{2}(-a_{i}^{(w_{0})+}-a_{i}^{(v_{0})+})}+e^{\frac{1}{2}(-a_{i}^{(w_{0})-}-a_{i}^{(v_{0})-})}\biggl( e^{\sqrt{a_{i}^{(w_{0})-}a_{i}^{(v_{0})-}}} -1\biggl)\biggl],
\end{eqnarray}
\end{widetext}
By combining Eqs.~\eqref{poisson}-\eqref{qwe}, we obtain the bound given by Eq.~\eqref{cotavariable_texto}.

A simpler bound given by Eq.~\eqref{cota_texto} arises by considering that the relative deviation parameters are all equal. It can be obtained directly from Eq.~\eqref{cotavariable_texto}.

It is also worth mentioning that one could use a more exhaustive approach by considering different records of settings for each of the two yields (or error probabilities) to be compared via the CS constraint. However, our numerical simulations suggest that the improvement achieved by doing so is not really significant, while the analysis is much more cumbersome to implement numerically.

If we focus our attention on nearest-neighbour intensity correlations only, and call $v_{0}=a$, $w_{0}=b$ and $v_{1}=c$, then from Eqs.~\eqref{cotavariable_texto} and~\eqref{cota_texto} respectively, one trivially obtains:
\begin{widetext}
\begin{equation}
\label{cotavariabele_1}
\left\langle\Psi_{b,c, Z, n} \mid \Psi_{a, c, Z, n}\right\rangle \geq
\sum_{a_{i}\in A} p_{a_{i}}  \biggl[e^{\frac{1}{2}(-a_{i}^{(b)+}-a_{i}^{(a)+})}+
e^{\frac{1}{2}(-a_{i}^{(b)-}-a_{i}^{(a)-})}\biggl( e^{\sqrt{a_{i}^{(b)-}a_{i}^{(a)-}}} -1\biggl)\biggl] \equiv \sqrt{\tau_{a, b, c, n}^{\text{$\xi=1$}}},
\end{equation}
\end{widetext}
and
\begin{eqnarray}
\label{cotafija}
\left\langle\Psi_{b,c, Z, n} \mid \Psi_{a, c, Z, n}\right\rangle&\geq& \left[1-\sum_{a_{i}\in A}p_{a_{i}}\left( e^{-a_{i}^{-}}-e^{-a_{i}^{+}} \right)\right]\nonumber \\
&\equiv&\sqrt{\tau_{a, b, c, n}^{\text{$\xi=1$}}}.
\end{eqnarray}

\subsection{Truncated normal model}\label{sec:methodsB}

Following Eq.~\eqref{poisson} and using the same notation as in the previous section for nearest-neighbour intensity correlations, we have that the bound for the different possible combinations of settings when the correlation function is a TG distribution is given by
\begin{equation}
\begin{aligned}
\label{taugauss}
&\left[\sum_{a_{k+1}\in A} p_{a_{k+1}}\sum_{n_{k+1}=0}^{\infty}\sqrt{p_{n_{k+1}}|_{a_{k+1},a}p_{n_{k+1}}|_{a_{k+1},b}}\right]^{2}\\
&\geq\left[\sum_{a_{k+1}\in A} p_{a_{k+1}}\sum_{n_{k+1}=0}^{n_{\text{cut}}}\sqrt{p_{n_{k+1}}|_{a_{k+1},a}p_{n_{k+1}}|_{a_{k+1},b}}\right]^{2} \\
&\equiv \tau_{a,b,c,n}^{\xi=1}.
\end{aligned}
\end{equation}

For instance, we have that
\begin{widetext}
\begin{equation}
p_{n_{k+1}}|_{a_{k+1}, a}=\int_{\lambda}^{\Lambda}\frac{e^{-\alpha_{k+1}} \alpha_{k+1}^{n_{k+1}}}{n_{k+1} !} 
\frac{\phi\left(\gamma_{a_{k+1}, a}, \sigma_{a_{k+1}, a}^{2} ; \alpha_{k+1}\right)}{\left[\Phi\left(\gamma_{a_{k+1}, a}, \sigma_{a_{k+1}, a}^{2} ; \Lambda\right)-\Phi\left(\gamma_{a_{k+1}, a},\sigma_{a_{k+1}, a}^{2} ;\lambda \right)\right]} d \alpha_{k+1},
\end{equation}
\end{widetext}
where $\sigma_{a_{k+1},a}$ and $\gamma_{a_{k+1},a}$ respectively are the parental standard deviation and the parental mean value of the random variable $\alpha_{k+1}$, given that $a_{k}=a$. Similarly, $\lambda$ and $\Lambda$ define the truncation intervals of the TG distribution.

\section{Reference values for the linearized Cauchy-Schwarz constraints}\label{appendix:values}

We now present the reference values $\tilde{y}_{n, a, c}$ and $\tilde{h}_{n, a, c}$. For the moment, we neglect the effect of the dark counts of Bob's detectors and the random assignments of the double clicks that he performs. This means that the possible genuine detection outcomes for an $n$-photon pulse emitted by Alice are “no click” (00), “error” (01), “no error” (10), and “double click” (11). Their probabilities are (see Fig.~\ref{fig:reference} below)
\begin{equation}
\begin{aligned}
&p_{00}=(1-\eta)^{n}, \\
&p_{01}=\left(\eta \sin ^{2} \delta_{\mathrm{A}}+1-\eta\right)^{n}-(1-\eta)^{n}, \\
&p_{10}=\left(\eta \cos ^{2} \delta_{\mathrm{A}}+1-\eta\right)^{n}-(1-\eta)^{n}, \\
&p_{11}=1-p_{00}-p_{01}-p_{10}.
\end{aligned}
\end{equation}
\begin{figure}[H]
\centering
\includegraphics [width=8.6cm, height=3.1cm] {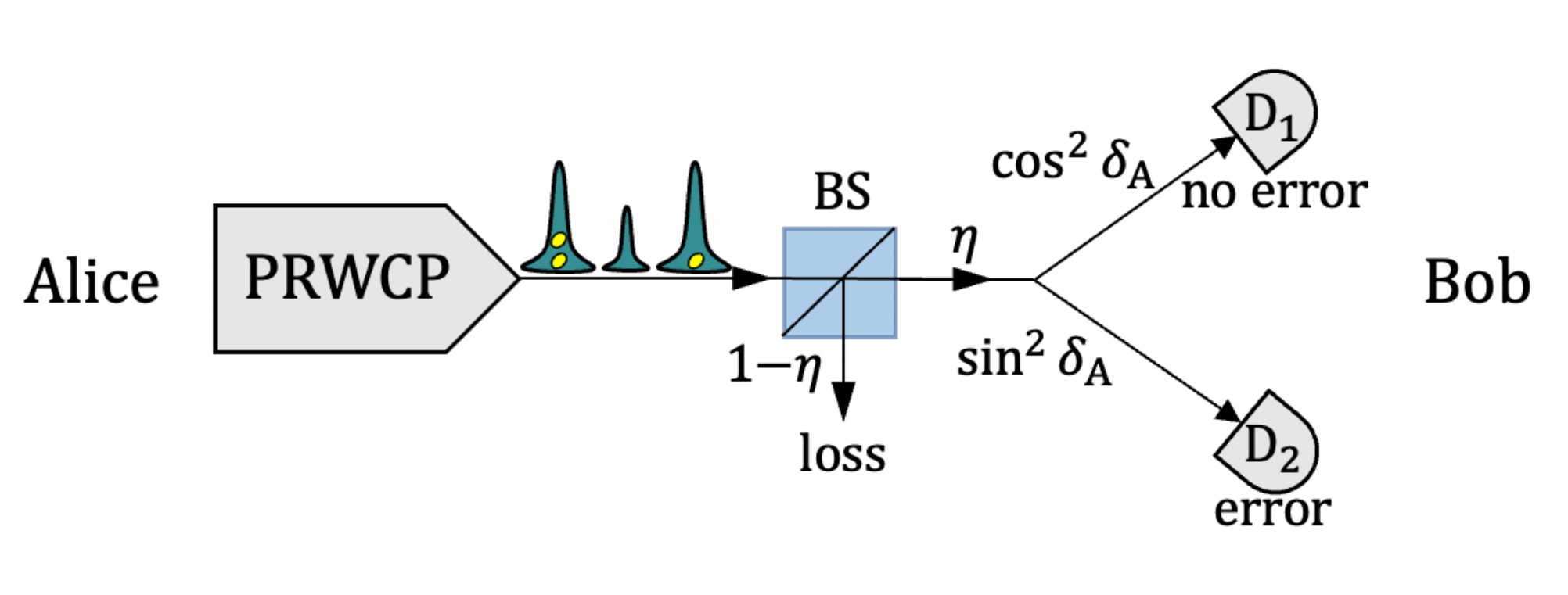}
\caption{Representation of the channel model considered for the simulations when Alice uses a phase-randomized weak coherent pulse (PRWCP) source. We recall that $\eta$ stands for the overall system efficiency ({\it i.e.}, it includes the loss in the channel and the finite detection efficiency of the detectors), and $\delta_{A}$ stands for the polarization misalignment introduced by the channel. We assume that Bob uses an active BB84 receiver with two detectors, $D_1$ and $D_2$. In the figure BS stands for beamsplitter. 
}
\label{fig:reference}
\end{figure}

We now need to incorporate the dark counts and the random assignments of the double clicks, so let us introduce the mutually exclusive events $A$ = \{no dark counts\}, $B$ = \{dark count in D1\}, $C$ = \{dark count in D2\} and $D$ = \{dark count in both D1 and D2\}, where we follow the detector notation of the figure above. The conditional error probabilities read
\begin{equation}
\begin{aligned}
&\left.p_{\mathrm{err}}\right|_{A}=p_{01}+\frac{1}{2} p_{11}, \\
&\left.p_{\mathrm{err}}\right|_{B}=\frac{1}{2}\left(p_{01}+p_{11}\right), \\
&\left.p_{\mathrm{err}}\right|_{C}=p_{00}+p_{01}+\frac{1}{2}\left(p_{10}+p_{11}\right), \\
&\left.p_{\mathrm{err}}\right|_{D}=\frac{1}{2},
\end{aligned}
\end{equation}
so that
\begin{eqnarray}
\tilde{h}_{n, a, c}&=&\left(1-p_{\mathrm{d}}\right)^{2} p_{\mathrm{err}}|_{A}+p_{\mathrm{d}}\left(1-p_{\mathrm{d}}\right)\left(\left.p_{\mathrm{err}}\right|_{B}+\left.p_{\mathrm{err}}\right|_{C}\right)\nonumber \\
&+&p_{\mathrm{d}}^{2} p_{\mathrm{err}}|_{D},
\end{eqnarray}
for all $n \in \mathbb{N}$ and $a, c \in A$. Notice that these quantities depend only on round $k$, so that they are independent of $c$ or the intensity setting of round $k-1$. Regarding the yield, we have:
\begin{equation}
\tilde{y}_{n, a, c}=1-\left(1-p_{\mathrm{d}}\right)^{2} p_{00}.
\end{equation}

\section{The truncated normal model}\label{appendix:truncated}
\begin{figure}[H]
\centering
\includegraphics [width=8.6cm, height=6cm] {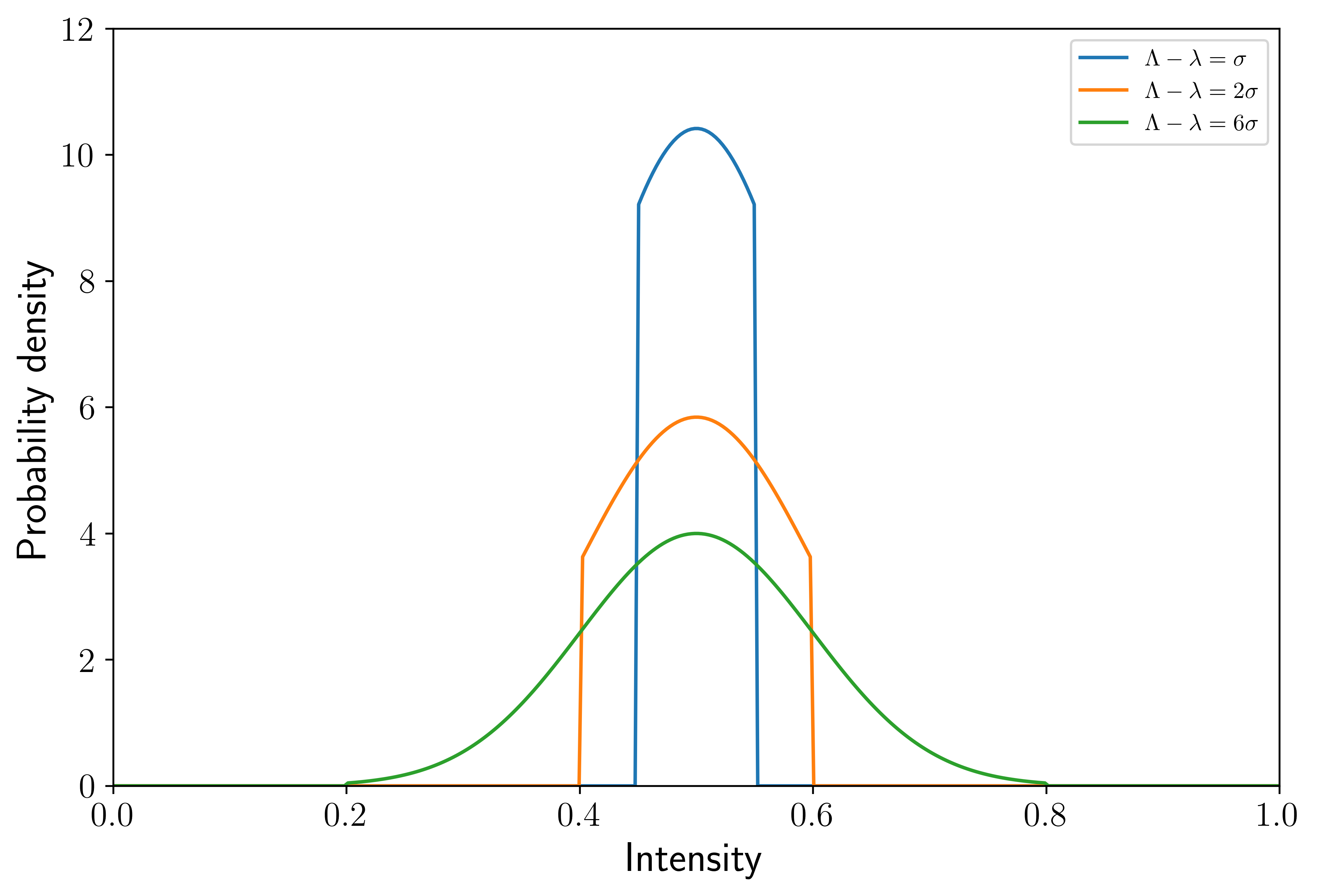}
\caption{Representation of a TG distribution with $\gamma=0.5$ and $\sigma=0.1$ for three different values of the truncation range. Shorter truncation intervals lead to a larger difference between $\sigma$ and $\tilde{\sigma}$. Note that, with $\Lambda-\lambda=6\sigma$, the gaussian shape of the distribution is preserved after truncation.}
\label{fig:truncated}
\end{figure}
\begin{figure}[H]
\includegraphics [width=8.6cm, height=6cm]{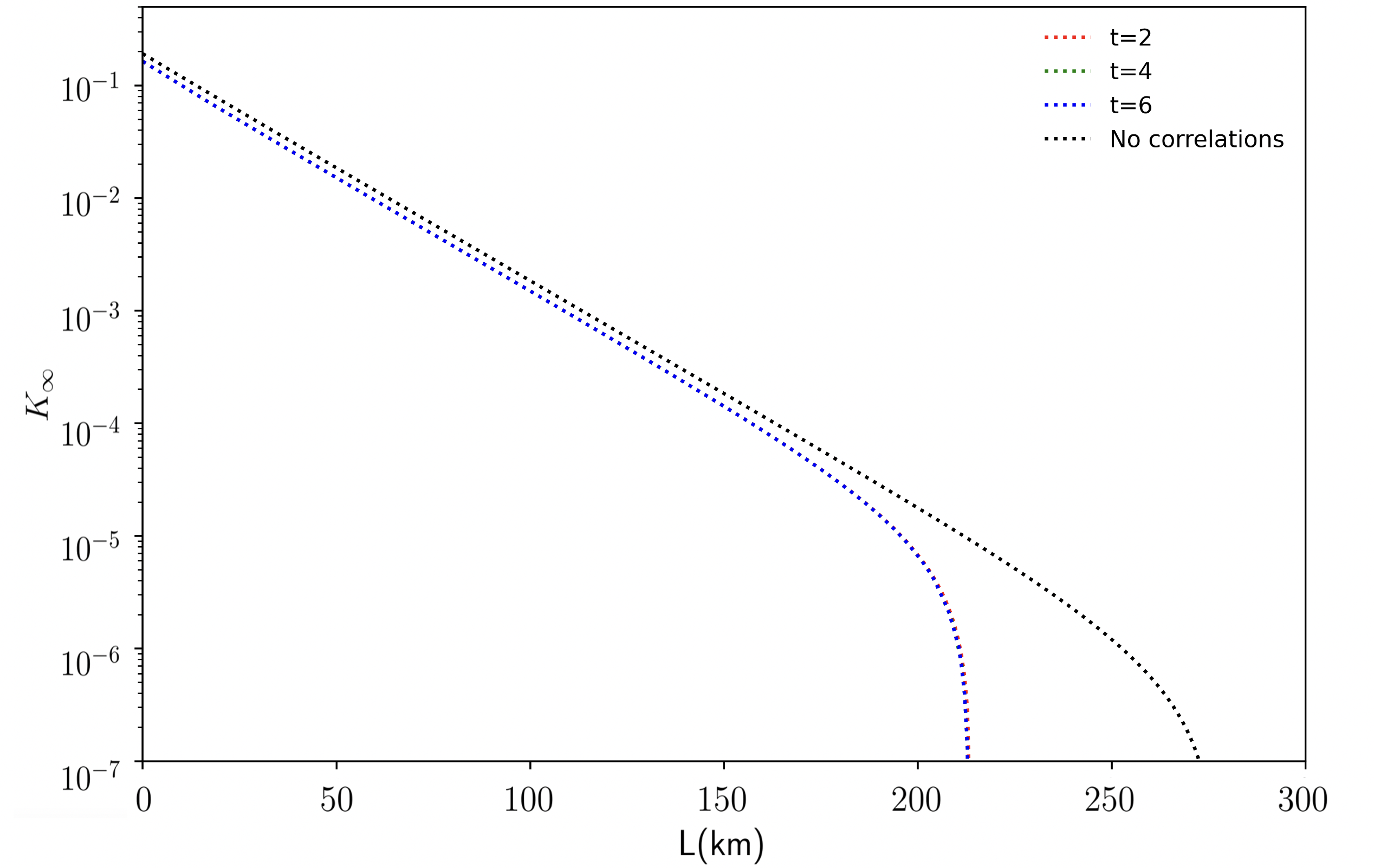} 
\caption{Effect of the parameter $t$ in the secret key rate corresponding to the TG model. The curves with $t=2,4$ and $6$ essentially overlap with each other. A bigger $t$ leads to a greater $\delta_{\text{max}}$ and as a consequence of that, the results corresponding to the model-independent case are worse. The three cases assume the same input values: $\tilde{\gamma}$, $\tilde{\sigma}$ and $[\lambda,\Lambda]$.}
\label{fig:t}
\end{figure}

In this Appendix, we present the link between the mean and variance of the parent normal distribution ($\gamma$ and $\sigma^{2}$), and the mean and variance of the truncated normal distribution ($\tilde{\gamma}$ and $\tilde{\sigma}^2$). For this purpose, we define
\begin{equation}
\alpha=\frac{\lambda-\gamma}{\sigma} \hspace{.2cm}\text{and}\hspace{.2cm} \beta=\frac{\Lambda-\gamma}{\sigma},
\end{equation}
so that we can make a mapping between the two \cite{truncated}, having:
\begin{eqnarray}
\label{mapping}
\tilde{\gamma}&=&\gamma-\sigma \frac{\phi(0,1 ; \beta)-\phi(0,1 ; \alpha)}{\Phi(0,1 ; \beta)-\Phi(0,1 ; \alpha)}\nonumber \\
\tilde{\sigma}^{2}&=&\sigma^{2} \biggl[1-\frac{\beta \phi(0,1 ; \beta)-\alpha \phi(0,1 ; \alpha)}{\Phi(0,1 ; \beta)-\Phi(0,1 ; \alpha)}\nonumber\\
&-&\left(\frac{\phi(0,1 ; \beta)-\phi(0,1 ; \alpha)}{\Phi(0,1 ; \beta)-\Phi(0,1 ; \alpha)}\right)^{2}\biggl],
\end{eqnarray}
where $\Phi(0,1 ; \alpha)$ and $\phi(0,1 ; \alpha)$ are defined in Eq.~\eqref{aux}.

To gain intuition about the form of this distribution and the effect that the truncation range has on $\tilde{\sigma}$, in Fig.~\ref{fig:truncated} we plot it for a couple of different intervals $[\lambda,\Lambda]$ centered in the mean value. Note that, in this way, the mean is not affected by the truncation.

To compare this model with the model-independent case, we need to relate the truncation interval and $\delta_{\text{max}}$, which we recall that fully characterizes the model independent setting. For this matter, we take $(\lambda,\Lambda)=(\tilde{\gamma} - t\tilde{\sigma}, \tilde{\gamma} + t\tilde{\sigma})$. We state the relation between $\delta_{\text{max}}$ and the parameter $t$ as $\tilde{\gamma}(1-\delta_{\text{max}})=\tilde{\gamma}-t\tilde{\sigma}$. In other words, we demand that the maximum relative deviation matches the extreme of the truncation. Note that, ideally, the interval $(\lambda,\Lambda)$ could be measured experimentally.

Regarding the value of $t$, for the sake of comparison with the model-independent case in the main text, we use $t=4$. It is also worth mentioning that modifying the value of $t$ has a negligible effect on the secret key rate $K_{\infty}$, as shown in Fig.~\ref{fig:t}.

\nocite{*}

\bibliography{aapmsamp}

\providecommand{\noopsort}[1]{}\providecommand{\singleletter}[1]{#1}%
\begin{thebibliography}{51}%
\makeatletter
\providecommand \@ifxundefined [1]{%
 \@ifx{#1\undefined}
}%
\providecommand \@ifnum [1]{%
 \ifnum #1\expandafter \@firstoftwo
 \else \expandafter \@secondoftwo
 \fi
}%
\providecommand \@ifx [1]{%
 \ifx #1\expandafter \@firstoftwo
 \else \expandafter \@secondoftwo
 \fi
}%
\providecommand \natexlab [1]{#1}%
\providecommand \enquote  [1]{``#1''}%
\providecommand \bibnamefont  [1]{#1}%
\providecommand \bibfnamefont [1]{#1}%
\providecommand \citenamefont [1]{#1}%
\providecommand \href@noop [0]{\@secondoftwo}%
\providecommand \href [0]{\begingroup \@sanitize@url \@href}%
\providecommand \@href[1]{\@@startlink{#1}\@@href}%
\providecommand \@@href[1]{\endgroup#1\@@endlink}%
\providecommand \@sanitize@url [0]{\catcode `\\12\catcode `\$12\catcode
  `\&12\catcode `\#12\catcode `\^12\catcode `\_12\catcode `\%12\relax}%
\providecommand \@@startlink[1]{}%
\providecommand \@@endlink[0]{}%
\providecommand \url  [0]{\begingroup\@sanitize@url \@url }%
\providecommand \@url [1]{\endgroup\@href {#1}{\urlprefix }}%
\providecommand \urlprefix  [0]{URL }%
\providecommand \Eprint [0]{\href }%
\providecommand \doibase [0]{https://doi.org/}%
\providecommand \selectlanguage [0]{\@gobble}%
\providecommand \bibinfo  [0]{\@secondoftwo}%
\providecommand \bibfield  [0]{\@secondoftwo}%
\providecommand \translation [1]{[#1]}%
\providecommand \BibitemOpen [0]{}%
\providecommand \bibitemStop [0]{}%
\providecommand \bibitemNoStop [0]{.\EOS\space}%
\providecommand \EOS [0]{\spacefactor3000\relax}%
\providecommand \BibitemShut  [1]{\csname bibitem#1\endcsname}%
\let\auto@bib@innerbib\@empty
\bibitem [{\citenamefont {Xu}\ \emph {et~al.}(2020)\citenamefont {Xu},
  \citenamefont {Ma}, \citenamefont {Zhang}, \citenamefont {Lo},\ and\
  \citenamefont {Pan}}]{extra1}%
  \BibitemOpen
  \bibfield  {author} {\bibinfo {author} {\bibfnamefont {F.}~\bibnamefont
  {Xu}}, \bibinfo {author} {\bibfnamefont {X.}~\bibnamefont {Ma}}, \bibinfo
  {author} {\bibfnamefont {Q.}~\bibnamefont {Zhang}}, \bibinfo {author}
  {\bibfnamefont {H.-K.}\ \bibnamefont {Lo}},\ and\ \bibinfo {author}
  {\bibfnamefont {J.-W.}\ \bibnamefont {Pan}},\ }\bibfield  {title} {\bibinfo
  {title} {Secure quantum key distribution with realistic devices},\ }\bibfield
   {journal} {\bibinfo  {journal} {Reviews of Modern Physics}\ }\textbf
  {\bibinfo {volume} {92}},\ \href
  {https://doi.org/10.1103/revmodphys.92.025002} {10.1103/revmodphys.92.025002}
  (\bibinfo {year} {2020})\BibitemShut {NoStop}%
\bibitem [{\citenamefont {Pirandola}\ \emph {et~al.}(2020)\citenamefont
  {Pirandola} \emph {et~al.}}]{extra2}%
  \BibitemOpen
  \bibfield  {author} {\bibinfo {author} {\bibfnamefont {S.}~\bibnamefont
  {Pirandola}} \emph {et~al.},\ }\bibfield  {title} {\bibinfo {title} {Advances
  in quantum cryptography},\ }\href {https://doi.org/10.1364/aop.361502}
  {\bibfield  {journal} {\bibinfo  {journal} {Advances in Optics and
  Photonics}\ }\textbf {\bibinfo {volume} {12}},\ \bibinfo {pages} {1012}
  (\bibinfo {year} {2020})}\BibitemShut {NoStop}%
\bibitem [{\citenamefont {Lo}\ \emph {et~al.}(2014)\citenamefont {Lo},
  \citenamefont {Curty},\ and\ \citenamefont {Tamaki}}]{extra3}%
  \BibitemOpen
  \bibfield  {author} {\bibinfo {author} {\bibfnamefont {H.-K.}\ \bibnamefont
  {Lo}}, \bibinfo {author} {\bibfnamefont {M.}~\bibnamefont {Curty}},\ and\
  \bibinfo {author} {\bibfnamefont {K.}~\bibnamefont {Tamaki}},\ }\bibfield
  {title} {\bibinfo {title} {Secure quantum key distribution},\ }\href
  {https://doi.org/10.1038/nphoton.2014.149} {\bibfield  {journal} {\bibinfo
  {journal} {Nature Photonics}\ }\textbf {\bibinfo {volume} {8}},\ \bibinfo
  {pages} {595} (\bibinfo {year} {2014})}\BibitemShut {NoStop}%
\bibitem [{\citenamefont {Vernam}(1926)}]{Vernam}%
  \BibitemOpen
  \bibfield  {author} {\bibinfo {author} {\bibfnamefont {G.}~\bibnamefont
  {Vernam}},\ }\bibfield  {journal} {\bibinfo  {journal} {Journal of the
  American Institute of Electrical Engineers}\ }\textbf {\bibinfo {volume}
  {45}},\ \href {https://doi.org/10.1109/T-AIEE.1926.5061224}
  {10.1109/T-AIEE.1926.5061224} (\bibinfo {year} {1926})\BibitemShut {NoStop}%
\bibitem [{\citenamefont {Sasaki}\ \emph {et~al.}(2011)\citenamefont {Sasaki}
  \emph {et~al.}}]{nueva1}%
  \BibitemOpen
  \bibfield  {author} {\bibinfo {author} {\bibfnamefont {M.}~\bibnamefont
  {Sasaki}} \emph {et~al.},\ }\bibfield  {title} {\bibinfo {title} {Field test
  of quantum key distribution in the tokyo qkd network},\ }\href
  {https://doi.org/10.1364/oe.19.010387} {\bibfield  {journal} {\bibinfo
  {journal} {Optics Express}\ }\textbf {\bibinfo {volume} {19}},\ \bibinfo
  {pages} {10387} (\bibinfo {year} {2011})}\BibitemShut {NoStop}%
\bibitem [{\citenamefont {Stucki}\ \emph {et~al.}(2011)\citenamefont {Stucki}
  \emph {et~al.}}]{nueva2}%
  \BibitemOpen
  \bibfield  {author} {\bibinfo {author} {\bibfnamefont {D.}~\bibnamefont
  {Stucki}} \emph {et~al.},\ }\bibfield  {title} {\bibinfo {title} {Long-term
  performance of the swissquantum quantum key distribution network in a field
  environment},\ }\href {https://doi.org/10.1088/1367-2630/13/12/123001}
  {\bibfield  {journal} {\bibinfo  {journal} {New Journal of Physics}\ }\textbf
  {\bibinfo {volume} {13}},\ \bibinfo {pages} {123001} (\bibinfo {year}
  {2011})}\BibitemShut {NoStop}%
\bibitem [{\citenamefont {Dynes}\ \emph {et~al.}(2019)\citenamefont {Dynes}
  \emph {et~al.}}]{nueva3}%
  \BibitemOpen
  \bibfield  {author} {\bibinfo {author} {\bibfnamefont {J.~F.}\ \bibnamefont
  {Dynes}} \emph {et~al.},\ }\bibfield  {title} {\bibinfo {title} {Cambridge
  quantum network},\ }\bibfield  {journal} {\bibinfo  {journal} {npj Quantum
  Information}\ }\textbf {\bibinfo {volume} {5}},\ \href
  {https://doi.org/10.1038/s41534-019-0221-4} {10.1038/s41534-019-0221-4}
  (\bibinfo {year} {2019})\BibitemShut {NoStop}%
\bibitem [{\citenamefont {Chen}\ \emph {et~al.}(2021)\citenamefont {Chen} \emph
  {et~al.}}]{nueva4}%
  \BibitemOpen
  \bibfield  {author} {\bibinfo {author} {\bibfnamefont {Y.-A.}\ \bibnamefont
  {Chen}} \emph {et~al.},\ }\bibfield  {title} {\bibinfo {title} {An integrated
  space-to-ground quantum communication network over 4,600 kilometres},\ }\href
  {https://doi.org/10.1038/s41586-020-03093-8} {\bibfield  {journal} {\bibinfo
  {journal} {Nature}\ }\textbf {\bibinfo {volume} {589}},\ \bibinfo {pages}
  {214} (\bibinfo {year} {2021})}\BibitemShut {NoStop}%
\bibitem [{\citenamefont {Liao}\ \emph {et~al.}(2017)\citenamefont {Liao} \emph
  {et~al.}}]{sat1}%
  \BibitemOpen
  \bibfield  {author} {\bibinfo {author} {\bibfnamefont {S.-K.}\ \bibnamefont
  {Liao}} \emph {et~al.},\ }\bibfield  {title} {\bibinfo {title}
  {Satellite-to-ground quantum key distribution},\ }\href
  {https://doi.org/10.1038/nature23655} {\bibfield  {journal} {\bibinfo
  {journal} {Nature}\ }\textbf {\bibinfo {volume} {549}},\ \bibinfo {pages}
  {43} (\bibinfo {year} {2017})}\BibitemShut {NoStop}%
\bibitem [{\citenamefont {Takenaka}\ \emph {et~al.}(2017)\citenamefont
  {Takenaka}, \citenamefont {Carrasco-Casado}, \citenamefont {Fujiwara},
  \citenamefont {Kitamura}, \citenamefont {Sasaki},\ and\ \citenamefont
  {Toyoshima}}]{sat2}%
  \BibitemOpen
  \bibfield  {author} {\bibinfo {author} {\bibfnamefont {H.}~\bibnamefont
  {Takenaka}}, \bibinfo {author} {\bibfnamefont {A.}~\bibnamefont
  {Carrasco-Casado}}, \bibinfo {author} {\bibfnamefont {M.}~\bibnamefont
  {Fujiwara}}, \bibinfo {author} {\bibfnamefont {M.}~\bibnamefont {Kitamura}},
  \bibinfo {author} {\bibfnamefont {M.}~\bibnamefont {Sasaki}},\ and\ \bibinfo
  {author} {\bibfnamefont {M.}~\bibnamefont {Toyoshima}},\ }\bibfield  {title}
  {\bibinfo {title} {Satellite-to-ground quantum-limited communication using a
  50-kg-class microsatellite},\ }\href
  {https://doi.org/10.1038/nphoton.2017.107} {\bibfield  {journal} {\bibinfo
  {journal} {Nature Photonics}\ }\textbf {\bibinfo {volume} {11}},\ \bibinfo
  {pages} {502} (\bibinfo {year} {2017})}\BibitemShut {NoStop}%
\bibitem [{\citenamefont {Liao}\ \emph {et~al.}(2018)\citenamefont {Liao} \emph
  {et~al.}}]{sat2b}%
  \BibitemOpen
  \bibfield  {author} {\bibinfo {author} {\bibfnamefont {S.-K.}\ \bibnamefont
  {Liao}} \emph {et~al.},\ }\bibfield  {title} {\bibinfo {title}
  {Satellite-relayed intercontinental quantum network},\ }\href
  {https://doi.org/10.1103/PhysRevLett.120.030501} {\bibfield  {journal}
  {\bibinfo  {journal} {Physical Review Letters}\ }\textbf {\bibinfo {volume}
  {120}},\ \bibinfo {pages} {030501} (\bibinfo {year} {2018})}\BibitemShut
  {NoStop}%
\bibitem [{\citenamefont {J.}\ \emph {et~al.}(2020)\citenamefont {J.} \emph
  {et~al.}}]{sat3}%
  \BibitemOpen
  \bibfield  {author} {\bibinfo {author} {\bibfnamefont {Y.}~\bibnamefont {J.}}
  \emph {et~al.},\ }\bibfield  {title} {\bibinfo {title} {Entanglement-based
  secure quantum cryptography over 1,120 kilometres},\ }\href
  {https://doi.org/10.1038/s41586-020-2401-y} {\bibfield  {journal} {\bibinfo
  {journal} {Nature}\ }\textbf {\bibinfo {volume} {582}},\ \bibinfo {pages}
  {501} (\bibinfo {year} {2020})}\BibitemShut {NoStop}%
\bibitem [{\citenamefont {Hwang}(2003)}]{decoy1}%
  \BibitemOpen
  \bibfield  {author} {\bibinfo {author} {\bibfnamefont {W.-Y.}\ \bibnamefont
  {Hwang}},\ }\bibfield  {title} {\bibinfo {title} {Quantum key distribution
  with high loss: Toward global secure communication},\ }\bibfield  {journal}
  {\bibinfo  {journal} {Physical Review Letters}\ }\textbf {\bibinfo {volume}
  {91}},\ \href {https://doi.org/10.1103/physrevlett.91.057901}
  {10.1103/physrevlett.91.057901} (\bibinfo {year} {2003})\BibitemShut
  {NoStop}%
\bibitem [{\citenamefont {Wang}(2005)}]{decoy2}%
  \BibitemOpen
  \bibfield  {author} {\bibinfo {author} {\bibfnamefont {X.-B.}\ \bibnamefont
  {Wang}},\ }\bibfield  {title} {\bibinfo {title} {Beating the
  photon-number-splitting attack in practical quantum cryptography},\
  }\bibfield  {journal} {\bibinfo  {journal} {Physical Review Letters}\
  }\textbf {\bibinfo {volume} {94}},\ \href
  {https://doi.org/10.1103/physrevlett.94.230503}
  {10.1103/physrevlett.94.230503} (\bibinfo {year} {2005})\BibitemShut
  {NoStop}%
\bibitem [{\citenamefont {Lo}\ \emph {et~al.}(2005)\citenamefont {Lo},
  \citenamefont {Ma},\ and\ \citenamefont {Chen}}]{Lo_2005}%
  \BibitemOpen
  \bibfield  {author} {\bibinfo {author} {\bibfnamefont {H.-K.}\ \bibnamefont
  {Lo}}, \bibinfo {author} {\bibfnamefont {X.}~\bibnamefont {Ma}},\ and\
  \bibinfo {author} {\bibfnamefont {K.}~\bibnamefont {Chen}},\ }\bibfield
  {title} {\bibinfo {title} {Decoy state quantum key distribution},\ }\bibfield
   {journal} {\bibinfo  {journal} {Physical Review Letters}\ }\textbf {\bibinfo
  {volume} {94}},\ \href {https://doi.org/10.1103/physrevlett.94.230504}
  {10.1103/physrevlett.94.230504} (\bibinfo {year} {2005})\BibitemShut
  {NoStop}%
\bibitem [{\citenamefont {Lim}\ \emph {et~al.}(2014)\citenamefont {Lim},
  \citenamefont {Curty}, \citenamefont {Walenta}, \citenamefont {Xu},\ and\
  \citenamefont {Zbinden}}]{security_decoy}%
  \BibitemOpen
  \bibfield  {author} {\bibinfo {author} {\bibfnamefont {C.~C.~W.}\
  \bibnamefont {Lim}}, \bibinfo {author} {\bibfnamefont {M.}~\bibnamefont
  {Curty}}, \bibinfo {author} {\bibfnamefont {N.}~\bibnamefont {Walenta}},
  \bibinfo {author} {\bibfnamefont {F.}~\bibnamefont {Xu}},\ and\ \bibinfo
  {author} {\bibfnamefont {H.}~\bibnamefont {Zbinden}},\ }\bibfield  {title}
  {\bibinfo {title} {Concise security bounds for practical decoy-state quantum
  key distribution},\ }\href {https://doi.org/10.1103/physreva.89.022307}
  {\bibfield  {journal} {\bibinfo  {journal} {Physical Review A}\ }\textbf
  {\bibinfo {volume} {89}},\ \bibinfo {pages} {022307} (\bibinfo {year}
  {2014})}\BibitemShut {NoStop}%
\bibitem [{\citenamefont {Zhao}\ \emph {et~al.}(2006)\citenamefont {Zhao} \emph
  {et~al.}}]{exp_decoy1}%
  \BibitemOpen
  \bibfield  {author} {\bibinfo {author} {\bibfnamefont {Y.}~\bibnamefont
  {Zhao}} \emph {et~al.},\ }\bibfield  {title} {\bibinfo {title} {Experimental
  quantum key distribution with decoy states},\ }\href
  {https://doi.org/10.1103/physreva.92.032305} {\bibfield  {journal} {\bibinfo
  {journal} {Physical Review Letters}\ }\textbf {\bibinfo {volume} {96}},\
  \bibinfo {pages} {70502} (\bibinfo {year} {2006})}\BibitemShut {NoStop}%
\bibitem [{\citenamefont {Yuan}\ \emph {et~al.}(2007)\citenamefont {Yuan} \emph
  {et~al.}}]{exp_decoy2}%
  \BibitemOpen
  \bibfield  {author} {\bibinfo {author} {\bibfnamefont {Z.~L.}\ \bibnamefont
  {Yuan}} \emph {et~al.},\ }\bibfield  {title} {\bibinfo {title}
  {Unconditionally secure one-way quantum key distribution using decoy
  pulses},\ }\href {https://doi.org/10.1063/1.2752766} {\bibfield  {journal}
  {\bibinfo  {journal} {Applied Physics Letters}\ }\textbf {\bibinfo {volume}
  {90}},\ \bibinfo {pages} {011118} (\bibinfo {year} {2007})}\BibitemShut
  {NoStop}%
\bibitem [{\citenamefont {Rosenberg}\ \emph {et~al.}(2007)\citenamefont
  {Rosenberg} \emph {et~al.}}]{exp_decoy3}%
  \BibitemOpen
  \bibfield  {author} {\bibinfo {author} {\bibfnamefont {D.}~\bibnamefont
  {Rosenberg}} \emph {et~al.},\ }\bibfield  {title} {\bibinfo {title}
  {Long-distance decoy-state quantum key distribution in optical fiber},\
  }\href {https://doi.org/10.1103/physrevlett.98.010503} {\bibfield  {journal}
  {\bibinfo  {journal} {Physical Review Letters}\ }\textbf {\bibinfo {volume}
  {98}},\ \bibinfo {pages} {10503} (\bibinfo {year} {2007})}\BibitemShut
  {NoStop}%
\bibitem [{\citenamefont {Peng}\ \emph {et~al.}(2007)\citenamefont {Peng} \emph
  {et~al.}}]{exp_decoy4}%
  \BibitemOpen
  \bibfield  {author} {\bibinfo {author} {\bibfnamefont {C.-Z.}\ \bibnamefont
  {Peng}} \emph {et~al.},\ }\bibfield  {title} {\bibinfo {title} {Experimental
  long-distance decoy-state quantum key distribution based on polarization
  encoding},\ }\href {https://doi.org/10.1103/physrevlett.98.010505} {\bibfield
   {journal} {\bibinfo  {journal} {Physical Review Letters}\ }\textbf {\bibinfo
  {volume} {98}},\ \bibinfo {pages} {010505} (\bibinfo {year}
  {2007})}\BibitemShut {NoStop}%
\bibitem [{\citenamefont {Schmitt-Manderbach}\ \emph
  {et~al.}(2007)\citenamefont {Schmitt-Manderbach} \emph
  {et~al.}}]{exp_decoy5}%
  \BibitemOpen
  \bibfield  {author} {\bibinfo {author} {\bibfnamefont {T.}~\bibnamefont
  {Schmitt-Manderbach}} \emph {et~al.},\ }\bibfield  {title} {\bibinfo {title}
  {Experimental demonstration of free-space decoy-state quantum key
  distribution over 144 km},\ }\href
  {https://doi.org/10.1103/PhysRevLett.98.010504} {\bibfield  {journal}
  {\bibinfo  {journal} {Physical Review Letters}\ }\textbf {\bibinfo {volume}
  {98}},\ \bibinfo {pages} {10504} (\bibinfo {year} {2007})}\BibitemShut
  {NoStop}%
\bibitem [{\citenamefont {Dixon}\ \emph {et~al.}(2008)\citenamefont {Dixon}
  \emph {et~al.}}]{exp_decoy5b}%
  \BibitemOpen
  \bibfield  {author} {\bibinfo {author} {\bibfnamefont {A.~R.}\ \bibnamefont
  {Dixon}} \emph {et~al.},\ }\bibfield  {title} {\bibinfo {title} {Gigahertz
  decoy quantum key distribution with 1 mbit/s secure key rate},\ }\href
  {https://doi.org/10.1364/oe.16.018790} {\bibfield  {journal} {\bibinfo
  {journal} {Optics Express}\ }\textbf {\bibinfo {volume} {16}},\ \bibinfo
  {pages} {18790} (\bibinfo {year} {2008})}\BibitemShut {NoStop}%
\bibitem [{\citenamefont {Liu}\ \emph {et~al.}(2010)\citenamefont {Liu} \emph
  {et~al.}}]{exp_decoy6}%
  \BibitemOpen
  \bibfield  {author} {\bibinfo {author} {\bibfnamefont {Y.}~\bibnamefont
  {Liu}} \emph {et~al.},\ }\bibfield  {title} {\bibinfo {title} {Decoy-state
  quantum key distribution with polarized photons over 200 km},\ }\href
  {https://doi.org/10.1364/OE.18.008587} {\bibfield  {journal} {\bibinfo
  {journal} {Optics Express}\ }\textbf {\bibinfo {volume} {18}},\ \bibinfo
  {pages} {8587} (\bibinfo {year} {2010})}\BibitemShut {NoStop}%
\bibitem [{\citenamefont {Frohlich}\ \emph {et~al.}(2017)\citenamefont
  {Frohlich} \emph {et~al.}}]{exp_decoy7}%
  \BibitemOpen
  \bibfield  {author} {\bibinfo {author} {\bibfnamefont {B.}~\bibnamefont
  {Frohlich}} \emph {et~al.},\ }\bibfield  {title} {\bibinfo {title}
  {Long-distance quantum key distribution secure against coherent attacks},\
  }\href {https://doi.org/10.1364/optica.4.000163} {\bibfield  {journal}
  {\bibinfo  {journal} {Optica}\ }\textbf {\bibinfo {volume} {4}},\ \bibinfo
  {pages} {163} (\bibinfo {year} {2017})}\BibitemShut {NoStop}%
\bibitem [{\citenamefont {Yuan}\ \emph {et~al.}(2018)\citenamefont {Yuan} \emph
  {et~al.}}]{exp_decoy8}%
  \BibitemOpen
  \bibfield  {author} {\bibinfo {author} {\bibfnamefont {Z.}~\bibnamefont
  {Yuan}} \emph {et~al.},\ }\bibfield  {title} {\bibinfo {title} {10-mb/s
  quantum key distribution},\ }\href {https://doi.org/10.1109/jlt.2018.2843136}
  {\bibfield  {journal} {\bibinfo  {journal} {J. Lightwave Tech.}\ }\textbf
  {\bibinfo {volume} {36}},\ \bibinfo {pages} {3427} (\bibinfo {year}
  {2018})}\BibitemShut {NoStop}%
\bibitem [{\citenamefont {Boaron}\ \emph
  {et~al.}(2018{\natexlab{a}})\citenamefont {Boaron} \emph
  {et~al.}}]{exp_decoy_long}%
  \BibitemOpen
  \bibfield  {author} {\bibinfo {author} {\bibfnamefont {A.}~\bibnamefont
  {Boaron}} \emph {et~al.},\ }\bibfield  {title} {\bibinfo {title} {Secure
  quantum key distribution over 421 km of optical fiber},\ }\href
  {https://doi.org/10.1103/physrevlett.121.190502} {\bibfield  {journal}
  {\bibinfo  {journal} {Physical Review Letters}\ }\textbf {\bibinfo {volume}
  {121}},\ \bibinfo {pages} {190502} (\bibinfo {year}
  {2018}{\natexlab{a}})}\BibitemShut {NoStop}%
\bibitem [{\citenamefont {Lo}\ \emph {et~al.}(2012)\citenamefont {Lo},
  \citenamefont {Curty},\ and\ \citenamefont {Qi}}]{mdi}%
  \BibitemOpen
  \bibfield  {author} {\bibinfo {author} {\bibfnamefont {H.-K.}\ \bibnamefont
  {Lo}}, \bibinfo {author} {\bibfnamefont {M.}~\bibnamefont {Curty}},\ and\
  \bibinfo {author} {\bibfnamefont {B.}~\bibnamefont {Qi}},\ }\bibfield
  {title} {\bibinfo {title} {Measurement-device-independent quantum key
  distribution},\ }\bibfield  {journal} {\bibinfo  {journal} {Physical Review
  Letters}\ }\textbf {\bibinfo {volume} {108}},\ \href
  {https://doi.org/10.1103/physrevlett.108.130503}
  {10.1103/physrevlett.108.130503} (\bibinfo {year} {2012})\BibitemShut
  {NoStop}%
\bibitem [{\citenamefont {Lucamarini}\ \emph
  {et~al.}(2018{\natexlab{a}})\citenamefont {Lucamarini}, \citenamefont {Yuan},
  \citenamefont {Dynes},\ and\ \citenamefont {Shields}}]{twinf}%
  \BibitemOpen
  \bibfield  {author} {\bibinfo {author} {\bibfnamefont {M.}~\bibnamefont
  {Lucamarini}}, \bibinfo {author} {\bibfnamefont {Z.~L.}\ \bibnamefont
  {Yuan}}, \bibinfo {author} {\bibfnamefont {J.~F.}\ \bibnamefont {Dynes}},\
  and\ \bibinfo {author} {\bibfnamefont {A.~J.}\ \bibnamefont {Shields}},\
  }\bibfield  {title} {\bibinfo {title} {Overcoming the rate-distance limit of
  quantum key distribution without quantum repeaters},\ }\href
  {https://doi.org/10.1038/s41586-018-0066-6} {\bibfield  {journal} {\bibinfo
  {journal} {Nature}\ }\textbf {\bibinfo {volume} {557}},\ \bibinfo {pages}
  {400} (\bibinfo {year} {2018}{\natexlab{a}})}\BibitemShut {NoStop}%
\bibitem [{\citenamefont {Wang}\ \emph {et~al.}(2022)\citenamefont {Wang} \emph
  {et~al.}}]{exp_tf}%
  \BibitemOpen
  \bibfield  {author} {\bibinfo {author} {\bibfnamefont {S.}~\bibnamefont
  {Wang}} \emph {et~al.},\ }\bibfield  {title} {\bibinfo {title} {Twin-field
  quantum key distribution over 830-km fibre},\ }\href
  {https://doi.org/10.1038/s41566-021-00928-2} {\bibfield  {journal} {\bibinfo
  {journal} {Nature Photonics}\ }\textbf {\bibinfo {volume} {16}},\ \bibinfo
  {pages} {154} (\bibinfo {year} {2022})}\BibitemShut {NoStop}%
\bibitem [{\citenamefont {Yoshino}\ \emph {et~al.}(2012)\citenamefont {Yoshino}
  \emph {et~al.}}]{multi1}%
  \BibitemOpen
  \bibfield  {author} {\bibinfo {author} {\bibfnamefont {K.~I.}\ \bibnamefont
  {Yoshino}} \emph {et~al.},\ }\bibfield  {title} {\bibinfo {title} {High-speed
  wavelength-division multiplexing quantum key distribution system},\ }\href
  {https://doi.org/10.1364/OL.37.000223} {\bibfield  {journal} {\bibinfo
  {journal} {Opt. Lett.}\ }\textbf {\bibinfo {volume} {37}},\ \bibinfo {pages}
  {223} (\bibinfo {year} {2012})}\BibitemShut {NoStop}%
\bibitem [{\citenamefont {Mora}\ \emph {et~al.}(2012)\citenamefont {Mora} \emph
  {et~al.}}]{multi2}%
  \BibitemOpen
  \bibfield  {author} {\bibinfo {author} {\bibfnamefont {J.}~\bibnamefont
  {Mora}} \emph {et~al.},\ }\bibfield  {title} {\bibinfo {title} {Simultaneous
  transmission of 20x2 wdm/scm-qkd and 4 bidirectional classical channels over
  a pon},\ }\href {https://doi.org/10.1364/OE.20.016358} {\bibfield  {journal}
  {\bibinfo  {journal} {Opt. Lett.}\ }\textbf {\bibinfo {volume} {20}},\
  \bibinfo {pages} {16358} (\bibinfo {year} {2012})}\BibitemShut {NoStop}%
\bibitem [{\citenamefont {Islam}\ \emph {et~al.}(2017)\citenamefont {Islam}
  \emph {et~al.}}]{HD_QKD}%
  \BibitemOpen
  \bibfield  {author} {\bibinfo {author} {\bibfnamefont {N.~T.}\ \bibnamefont
  {Islam}} \emph {et~al.},\ }\bibfield  {title} {\bibinfo {title}
  {Provably-secure and high-rate quantum key distribution with time-bin
  qudits},\ }\href {https://doi.org/10.1126/sciadv.1701491} {\bibfield
  {journal} {\bibinfo  {journal} {Science Advances}\ }\textbf {\bibinfo
  {volume} {3}},\ \bibinfo {pages} {e1701491} (\bibinfo {year}
  {2017})}\BibitemShut {NoStop}%
\bibitem [{\citenamefont {Grünenfelder}\ \emph {et~al.}(2020)\citenamefont
  {Grünenfelder}, \citenamefont {Boaron}, \citenamefont {Rusca}, \citenamefont
  {Martin},\ and\ \citenamefont {Zbinden}}]{Fadri}%
  \BibitemOpen
  \bibfield  {author} {\bibinfo {author} {\bibfnamefont {F.}~\bibnamefont
  {Grünenfelder}}, \bibinfo {author} {\bibfnamefont {A.}~\bibnamefont
  {Boaron}}, \bibinfo {author} {\bibfnamefont {D.}~\bibnamefont {Rusca}},
  \bibinfo {author} {\bibfnamefont {A.}~\bibnamefont {Martin}},\ and\ \bibinfo
  {author} {\bibfnamefont {H.}~\bibnamefont {Zbinden}},\ }\bibfield  {title}
  {\bibinfo {title} {Performance and security of 5ghz repetition rate
  polarization-based quantum key distribution},\ }\href
  {https://doi.org/10.1063/5.0021468} {\bibfield  {journal} {\bibinfo
  {journal} {Applied Physics Letters}\ }\textbf {\bibinfo {volume} {117}},\
  \bibinfo {pages} {144003} (\bibinfo {year} {2020})}\BibitemShut {NoStop}%
\bibitem [{\citenamefont {Yoshino}\ \emph {et~al.}(2018)\citenamefont {Yoshino}
  \emph {et~al.}}]{japoneses}%
  \BibitemOpen
  \bibfield  {author} {\bibinfo {author} {\bibfnamefont {K.-i.}\ \bibnamefont
  {Yoshino}} \emph {et~al.},\ }\bibfield  {title} {\bibinfo {title} {Quantum
  key distribution with an efficient countermeasure against correlated
  intensity fluctuations in optical pulses},\ }\bibfield  {journal} {\bibinfo
  {journal} {npj Quantum Information}\ }\textbf {\bibinfo {volume} {4}},\ \href
  {https://doi.org/10.1038/s41534-017-0057-8} {10.1038/s41534-017-0057-8}
  (\bibinfo {year} {2018})\BibitemShut {NoStop}%
\bibitem [{\citenamefont {Pereira}\ \emph {et~al.}(2020)\citenamefont
  {Pereira}, \citenamefont {Kato}, \citenamefont {Mizutani}, \citenamefont
  {Curty},\ and\ \citenamefont {Tamaki}}]{QKD_correlated}%
  \BibitemOpen
  \bibfield  {author} {\bibinfo {author} {\bibfnamefont {M.}~\bibnamefont
  {Pereira}}, \bibinfo {author} {\bibfnamefont {G.}~\bibnamefont {Kato}},
  \bibinfo {author} {\bibfnamefont {A.}~\bibnamefont {Mizutani}}, \bibinfo
  {author} {\bibfnamefont {M.}~\bibnamefont {Curty}},\ and\ \bibinfo {author}
  {\bibfnamefont {K.}~\bibnamefont {Tamaki}},\ }\bibfield  {title} {\bibinfo
  {title} {Quantum key distribution with correlated sources},\ }\bibfield
  {journal} {\bibinfo  {journal} {Science Advances}\ }\textbf {\bibinfo
  {volume} {6}},\ \href {https://doi.org/10.1126/sciadv.aaz4487}
  {10.1126/sciadv.aaz4487} (\bibinfo {year} {2020})\BibitemShut {NoStop}%
\bibitem [{\citenamefont {Nagamatsu}\ \emph {et~al.}(2016)\citenamefont
  {Nagamatsu}, \citenamefont {Mizutani}, \citenamefont {Ikuta}, \citenamefont
  {Yamamoto}, \citenamefont {Imoto},\ and\ \citenamefont {Tamaki}}]{corr1}%
  \BibitemOpen
  \bibfield  {author} {\bibinfo {author} {\bibfnamefont {Y.}~\bibnamefont
  {Nagamatsu}}, \bibinfo {author} {\bibfnamefont {A.}~\bibnamefont {Mizutani}},
  \bibinfo {author} {\bibfnamefont {R.}~\bibnamefont {Ikuta}}, \bibinfo
  {author} {\bibfnamefont {T.}~\bibnamefont {Yamamoto}}, \bibinfo {author}
  {\bibfnamefont {N.}~\bibnamefont {Imoto}},\ and\ \bibinfo {author}
  {\bibfnamefont {K.}~\bibnamefont {Tamaki}},\ }\bibfield  {title} {\bibinfo
  {title} {Security of quantum key distribution with light sources that are not
  independently and identically distributed},\ }\href
  {https://doi.org/10.1103/physreva.93.042325} {\bibfield  {journal} {\bibinfo
  {journal} {Phys. Rev. A}\ }\textbf {\bibinfo {volume} {93}},\ \bibinfo
  {pages} {042325} (\bibinfo {year} {2016})}\BibitemShut {NoStop}%
\bibitem [{\citenamefont {Mizutani}\ \emph {et~al.}(2019)\citenamefont
  {Mizutani}, \citenamefont {Kato}, \citenamefont {Azuma}, \citenamefont
  {Curty}, \citenamefont {Ikuta}, \citenamefont {Yamamoto}, \citenamefont
  {Imoto}, \citenamefont {Lo},\ and\ \citenamefont {Tamaki}}]{corr2}%
  \BibitemOpen
  \bibfield  {author} {\bibinfo {author} {\bibfnamefont {A.}~\bibnamefont
  {Mizutani}}, \bibinfo {author} {\bibfnamefont {G.}~\bibnamefont {Kato}},
  \bibinfo {author} {\bibfnamefont {K.}~\bibnamefont {Azuma}}, \bibinfo
  {author} {\bibfnamefont {M.}~\bibnamefont {Curty}}, \bibinfo {author}
  {\bibfnamefont {R.}~\bibnamefont {Ikuta}}, \bibinfo {author} {\bibfnamefont
  {T.}~\bibnamefont {Yamamoto}}, \bibinfo {author} {\bibfnamefont
  {N.}~\bibnamefont {Imoto}}, \bibinfo {author} {\bibfnamefont {H.-K.}\
  \bibnamefont {Lo}},\ and\ \bibinfo {author} {\bibfnamefont {K.}~\bibnamefont
  {Tamaki}},\ }\bibfield  {title} {\bibinfo {title} {Quantum key distribution
  with setting-choice-independently correlated light sources},\ }\href
  {https://doi.org/10.1038/s41534-018-0122-y} {\bibfield  {journal} {\bibinfo
  {journal} {npj Quantum Inf.}\ }\textbf {\bibinfo {volume} {5}},\ \bibinfo
  {pages} {8} (\bibinfo {year} {2019})}\BibitemShut {NoStop}%
\bibitem [{\citenamefont {Zapatero}\ \emph {et~al.}(2021)\citenamefont
  {Zapatero}, \citenamefont {Navarrete}, \citenamefont {Tamaki},\ and\
  \citenamefont {Curty}}]{Zapatero}%
  \BibitemOpen
  \bibfield  {author} {\bibinfo {author} {\bibfnamefont {V.}~\bibnamefont
  {Zapatero}}, \bibinfo {author} {\bibfnamefont {A.}~\bibnamefont {Navarrete}},
  \bibinfo {author} {\bibfnamefont {K.}~\bibnamefont {Tamaki}},\ and\ \bibinfo
  {author} {\bibfnamefont {M.}~\bibnamefont {Curty}},\ }\bibfield  {title}
  {\bibinfo {title} {Security of quantum key distribution with intensity
  correlations},\ }\href {https://doi.org/10.22331/q-2021-12-07-602} {\bibfield
   {journal} {\bibinfo  {journal} {Quantum}\ }\textbf {\bibinfo {volume} {5}},\
  \bibinfo {pages} {602} (\bibinfo {year} {2021})}\BibitemShut {NoStop}%
\bibitem [{\citenamefont {Lo}\ and\ \citenamefont {Preskill}(2007)}]{lo}%
  \BibitemOpen
  \bibfield  {author} {\bibinfo {author} {\bibfnamefont {H.-K.}\ \bibnamefont
  {Lo}}\ and\ \bibinfo {author} {\bibfnamefont {J.}~\bibnamefont {Preskill}},\
  }\bibfield  {title} {\bibinfo {title} {Security of quantum key distribution
  using weak coherent states with nonrandom phases},\ }\href
  {https://doi.org/10.48550/ARXIV.QUANT-PH/0610203} {\bibfield  {journal}
  {\bibinfo  {journal} {Quantum Info. Comput.}\ }\textbf {\bibinfo {volume}
  {7}},\ \bibinfo {pages} {431–458} (\bibinfo {year} {2007})}\BibitemShut
  {NoStop}%
\bibitem [{\citenamefont {Nahar}\ and\ \citenamefont
  {Lütkenhaus}(2021)}]{corr3}%
  \BibitemOpen
  \bibfield  {author} {\bibinfo {author} {\bibfnamefont {S.}~\bibnamefont
  {Nahar}}\ and\ \bibinfo {author} {\bibfnamefont {N.}~\bibnamefont
  {Lütkenhaus}},\ }\bibfield  {title} {\bibinfo {title} {Quantum key
  distribution with characterized source defects},\ }in\ \href
  {https://2021.qcrypt.net/posters/QCrypt2021Poster215Nahar.pdf} {\emph
  {\bibinfo {booktitle} {Poster presented at the International Conference on
  Quantum Cryptography (QCRYPT)}}}\ (\bibinfo {year} {2021})\BibitemShut
  {NoStop}%
\bibitem [{\citenamefont {Lucamarini}\ \emph {et~al.}(2015)\citenamefont
  {Lucamarini}, \citenamefont {Choi}, \citenamefont {Ward}, \citenamefont
  {Dynes}, \citenamefont {Yuan},\ and\ \citenamefont {Shields}}]{tha1}%
  \BibitemOpen
  \bibfield  {author} {\bibinfo {author} {\bibfnamefont {M.}~\bibnamefont
  {Lucamarini}}, \bibinfo {author} {\bibfnamefont {I.}~\bibnamefont {Choi}},
  \bibinfo {author} {\bibfnamefont {M.~B.}\ \bibnamefont {Ward}}, \bibinfo
  {author} {\bibfnamefont {J.~F.}\ \bibnamefont {Dynes}}, \bibinfo {author}
  {\bibfnamefont {Z.}~\bibnamefont {Yuan}},\ and\ \bibinfo {author}
  {\bibfnamefont {A.~J.}\ \bibnamefont {Shields}},\ }\bibfield  {title}
  {\bibinfo {title} {Practical security bounds against the trojan-horse attack
  in quantum key distribution},\ }\href
  {https://doi.org/10.1103/physrevx.5.031030} {\bibfield  {journal} {\bibinfo
  {journal} {Physical Review X}\ }\textbf {\bibinfo {volume} {5}},\ \bibinfo
  {pages} {031030} (\bibinfo {year} {2015})}\BibitemShut {NoStop}%
\bibitem [{\citenamefont {Tamaki}\ \emph {et~al.}(2016)\citenamefont {Tamaki},
  \citenamefont {Curty},\ and\ \citenamefont {Lucamarini}}]{tha2}%
  \BibitemOpen
  \bibfield  {author} {\bibinfo {author} {\bibfnamefont {K.}~\bibnamefont
  {Tamaki}}, \bibinfo {author} {\bibfnamefont {M.}~\bibnamefont {Curty}},\ and\
  \bibinfo {author} {\bibfnamefont {M.}~\bibnamefont {Lucamarini}},\ }\bibfield
   {title} {\bibinfo {title} {Decoy-state quantum key distribution with a leaky
  source},\ }\href {https://doi.org/10.1088/1367-2630/18/6/065008} {\bibfield
  {journal} {\bibinfo  {journal} {New Journal of Physics}\ }\textbf {\bibinfo
  {volume} {18}},\ \bibinfo {pages} {065008} (\bibinfo {year}
  {2016})}\BibitemShut {NoStop}%
\bibitem [{\citenamefont {Navarrete}\ and\ \citenamefont {Curty}(2022)}]{tha3}%
  \BibitemOpen
  \bibfield  {author} {\bibinfo {author} {\bibfnamefont {A.}~\bibnamefont
  {Navarrete}}\ and\ \bibinfo {author} {\bibfnamefont {M.}~\bibnamefont
  {Curty}},\ }\bibfield  {title} {\bibinfo {title} {Improved finite-key
  security analysis of quantum key distribution against trojan-horse attacks},\
  }\href {https://doi.org/10.48550/ARXIV.2202.06630} {\bibfield  {journal}
  {\bibinfo  {journal} {Quantum Science and Technology}\ }\textbf {\bibinfo
  {volume} {7}},\ \bibinfo {pages} {035021} (\bibinfo {year}
  {2022})}\BibitemShut {NoStop}%
\bibitem [{\citenamefont {Huang}\ \emph {et~al.}(2022)\citenamefont {Huang},
  \citenamefont {Mizutani}, \citenamefont {Lo},\ and\ \citenamefont
  {Makarov}}]{maka}%
  \BibitemOpen
  \bibfield  {author} {\bibinfo {author} {\bibfnamefont {A.}~\bibnamefont
  {Huang}}, \bibinfo {author} {\bibfnamefont {A.}~\bibnamefont {Mizutani}},
  \bibinfo {author} {\bibfnamefont {H.-K.}\ \bibnamefont {Lo}},\ and\ \bibinfo
  {author} {\bibfnamefont {K.}~\bibnamefont {Makarov}, \bibfnamefont
  {V.and~Tamaki}},\ }\bibfield  {title} {\bibinfo {title} {Characterisation of
  state preparation uncertainty in quantum key distribution},\ }\bibfield
  {journal} {\bibinfo  {journal} {Preprint arXiv:2205.11870}\ }\href
  {https://doi.org/10.48550/ARXIV.2205.11870} {10.48550/ARXIV.2205.11870}
  (\bibinfo {year} {2022})\BibitemShut {NoStop}%
\bibitem [{\citenamefont {Christandl}\ \emph {et~al.}(2009)\citenamefont
  {Christandl}, \citenamefont {König},\ and\ \citenamefont {Renner}}]{extra4}%
  \BibitemOpen
  \bibfield  {author} {\bibinfo {author} {\bibfnamefont {M.}~\bibnamefont
  {Christandl}}, \bibinfo {author} {\bibfnamefont {R.}~\bibnamefont {König}},\
  and\ \bibinfo {author} {\bibfnamefont {R.}~\bibnamefont {Renner}},\
  }\bibfield  {title} {\bibinfo {title} {Postselection technique for quantum
  channels with applications to quantum cryptography},\ }\bibfield  {journal}
  {\bibinfo  {journal} {Physical Review Letters}\ }\textbf {\bibinfo {volume}
  {102}},\ \href {https://doi.org/10.1103/physrevlett.102.020504}
  {10.1103/physrevlett.102.020504} (\bibinfo {year} {2009})\BibitemShut
  {NoStop}%
\bibitem [{\citenamefont {Renner}(2007)}]{Finetti}%
  \BibitemOpen
  \bibfield  {author} {\bibinfo {author} {\bibfnamefont {R.}~\bibnamefont
  {Renner}},\ }\bibfield  {title} {\bibinfo {title} {Symmetry of large physical
  systems implies independence of subsystems},\ }\href
  {https://doi.org/10.1038/nphys684} {\bibfield  {journal} {\bibinfo  {journal}
  {Nature Physics}\ }\textbf {\bibinfo {volume} {3}},\ \bibinfo {pages}
  {645–649} (\bibinfo {year} {2007})}\BibitemShut {NoStop}%
\bibitem [{\citenamefont {Renner}(2008)}]{extra5}%
  \BibitemOpen
  \bibfield  {author} {\bibinfo {author} {\bibfnamefont {R.}~\bibnamefont
  {Renner}},\ }\bibfield  {title} {\bibinfo {title} {Security of quantum key
  distribution},\ }\href {https://doi.org/10.48550/ARXIV.QUANT-PH/0512258}
  {\bibfield  {journal} {\bibinfo  {journal} {International Journal of Quantum
  Information}\ }\textbf {\bibinfo {volume} {6}},\ \bibinfo {pages} {1}
  (\bibinfo {year} {2008})}\BibitemShut {NoStop}%
\bibitem [{\citenamefont {Yin}\ \emph {et~al.}(2016)\citenamefont {Yin} \emph
  {et~al.}}]{decrease1}%
  \BibitemOpen
  \bibfield  {author} {\bibinfo {author} {\bibfnamefont {H.-L.}\ \bibnamefont
  {Yin}} \emph {et~al.},\ }\bibfield  {title} {\bibinfo {title}
  {Measurement-device-independent quantum key distribution over a 404~km
  optical fiber},\ }\bibfield  {journal} {\bibinfo  {journal} {Physical Review
  Letters}\ }\textbf {\bibinfo {volume} {117}},\ \href
  {https://doi.org/10.1103/physrevlett.117.190501}
  {10.1103/physrevlett.117.190501} (\bibinfo {year} {2016})\BibitemShut
  {NoStop}%
\bibitem [{\citenamefont {Johnson}\ \emph {et~al.}(1994)\citenamefont
  {Johnson}, \citenamefont {Kotz},\ and\ \citenamefont
  {Balakrishnan}}]{truncated}%
  \BibitemOpen
  \bibfield  {author} {\bibinfo {author} {\bibfnamefont {N.}~\bibnamefont
  {Johnson}}, \bibinfo {author} {\bibfnamefont {S.}~\bibnamefont {Kotz}},\ and\
  \bibinfo {author} {\bibfnamefont {N.}~\bibnamefont {Balakrishnan}},\
  }\href@noop {} {\emph {\bibinfo {title} {Continuous Univariate Distributions,
  Volume 1}}},\ \bibinfo {edition} {1st}\ ed.\ (\bibinfo  {publisher}
  {Wiley-Interscience},\ \bibinfo {year} {1994})\BibitemShut {NoStop}%
\bibitem [{\citenamefont {Lucamarini}\ \emph
  {et~al.}(2018{\natexlab{b}})\citenamefont {Lucamarini}, \citenamefont {Yuan},
  \citenamefont {Dynes},\ and\ \citenamefont {Shields}}]{twin}%
  \BibitemOpen
  \bibfield  {author} {\bibinfo {author} {\bibfnamefont {M.}~\bibnamefont
  {Lucamarini}}, \bibinfo {author} {\bibfnamefont {Z.~L.}\ \bibnamefont
  {Yuan}}, \bibinfo {author} {\bibfnamefont {J.~F.}\ \bibnamefont {Dynes}},\
  and\ \bibinfo {author} {\bibfnamefont {A.~J.}\ \bibnamefont {Shields}},\
  }\bibfield  {title} {\bibinfo {title} {Overcoming the rate–distance limit
  of quantum key distribution without quantum repeaters},\ }\href
  {https://doi.org/10.1038/s41586-018-0066-6} {\bibfield  {journal} {\bibinfo
  {journal} {Nature}\ }\textbf {\bibinfo {volume} {557}},\ \bibinfo {pages}
  {400–403} (\bibinfo {year} {2018}{\natexlab{b}})}\BibitemShut {NoStop}%
\bibitem [{\citenamefont {Boaron}\ \emph
  {et~al.}(2018{\natexlab{b}})\citenamefont {Boaron} \emph
  {et~al.}}]{decrease2}%
  \BibitemOpen
  \bibfield  {author} {\bibinfo {author} {\bibfnamefont {A.}~\bibnamefont
  {Boaron}} \emph {et~al.},\ }\bibfield  {title} {\bibinfo {title} {Secure
  quantum key distribution over 421~km of optical fiber},\ }\bibfield
  {journal} {\bibinfo  {journal} {Physical Review Letters}\ }\textbf {\bibinfo
  {volume} {121}},\ \href {https://doi.org/10.1103/physrevlett.121.190502}
  {10.1103/physrevlett.121.190502} (\bibinfo {year}
  {2018}{\natexlab{b}})\BibitemShut {NoStop}%
\end{thebibliography}%

\end{document}